\documentclass[aps,prl,twocolumn,superscriptaddress,amsmath,amssymb,longbibliography]{revtex4-2}
\usepackage{graphicx}
\usepackage{bm}
\usepackage{xcolor}
\usepackage[colorlinks=true,linkcolor=blue,citecolor=blue,urlcolor=blue]{hyperref}

\newcommand{\ad}{\hat a^\dagger}
\newcommand{\aop}{\hat a}

\begin{document}

\title{Squeezed terahertz light from nonlinear diamagnetism in ultrastrongly coupled two-dimensional electrons}

\author{Dasom Kim}
\affiliation{Applied Physics Graduate Program, Smalley--Curl Institute, Rice University, Houston, TX 77005, USA}
\affiliation{Department of Electrical and Computer Engineering, Rice University, Houston, TX 77005, USA}
\author{Junichiro Kono}
\affiliation{Department of Electrical and Computer Engineering, Rice University, Houston, TX 77005, USA}
\affiliation{Department of Physics and Astronomy, Rice University, Houston, TX 77005, USA}
\affiliation{Smalley--Curl Institute, Rice University, Houston, TX 77005, USA}
\affiliation{Rice Advanced Materials Institute, Rice University, Houston, TX 77005, USA}
\affiliation{Department of Materials Science and NanoEngineering, Rice University, Houston, TX 77005, USA}
 \author{Stephen Hughes}
 \affiliation{Department of Physics, Engineering Physics and Astronomy, Queen's University, Kingston, Ontario K7L 3N6, Canada}
\date{\today}

\begin{abstract}
Two-dimensional electron gases in terahertz cavities are among the most mature platforms for ultrastrong light--matter coupling, yet their collective response is captured by the linear Hopfield model, which has an exact classical correspondence. We show that the intrinsic band nonparabolicity of the electrons, recently revealed through nonlinear diamagnetism, turns this system into a quantum Kerr oscillator that squeezes terahertz light. Using a non-rotating-wave theory with ultrastrong-coupling-consistent input--output, an ab initio cavity bath, and a self-consistent model of electron heating, we predict output squeezing of 1.8--2.4\,dB with present GaAs samples in weakly confined cavities with quality factors of order $10^2$, and up to $\approx9$\,dB with modest improvements in mobility or cavity frequency.
Carrier heating sets limits through a material figure of merit, which yields a simple design rule that the optimal cavity quality factor scales inversely with the ultrastrong-coupling strength.

\end{abstract}

\maketitle


The ultrastrong-coupling (USC) regime, in which the light--matter coupling becomes a sizable fraction of the bare excitation frequencies, has developed into a broad field at the interface of quantum optics and condensed matter physics~\cite{FriskKockum2019,FornDiaz2019}.
One of its most mature realizations is a two-dimensional electron gas (2DEG) in a terahertz (THz) cavity~\cite{Scalari2012,Mavrona2021,Tay2025}, where normalized couplings beyond unity~\cite{Bayer2017}, record cooperativities~\cite{Zhang2016} and the vacuum Bloch--Siegert shift~\cite{Li2018} have been demonstrated. Superconducting circuits, by contrast, realize the quantum Rabi model with single artificial atoms~\cite{Niemczyk2010,Yoshihara2017}.

For collective systems, the polariton spectra are well described by the linear Hopfield model~\cite{Hopfield1958,Ciuti2005}. Nonlinear USC experiments have also been reported, including subcycle nonlinearities and wave mixing of Landau polaritons~\cite{Halbhuber2020,Mornhinweg2021}, alongside nonlinear exciton and intersubband polaritons (see Ref.~\cite{Kim2026} and references therein), but these are classical nonlinear responses of the coupled system rather than nonclassical states of the output field. Being quadratic in the bosonic operators, this model has an exact classical correspondence under classical driving, so its spectra follow from classical linear optics once the cavity--bath coupling and the gauge are treated consistently~\cite{DeBernardis2018,DiStefano2019,Salmon2022,Hughes2024,Gustin2025}. Thus, despite the striking USC phenomena observed with 2DEGs, including the vacuum Bloch--Siegert shift~\cite{Li2018}, the input--output experiments used to probe polaritonic features have so far revealed no genuinely quantum-optical field effects.

Proposed signatures of the squeezed USC vacuum include virtual photons that are not radiated~\cite{Ciuti2005}. Converting them into real, nonclassical photons requires a parametric process. 
One route is an external modulation of the coupling at the sum of two polariton frequencies, the dynamical Casimir effect for a time-modulated vacuum Rabi frequency~\cite{DeLiberato2007,Akbari2025}; for Landau polaritons, modulating the magnetic field produces dynamical Casimir pair emission with up to 5--10\,dB of two-mode squeezing~\cite{Hughes2026Floquet}. The other is a nonlinearity, which turns a simple monochromatic drive into an internal parametric pump for the quantum fluctuations.

A recent experiment by Kim \textit{et al.}~\cite{Kim2026} showed that such a nonlinearity exists in precisely the GaAs 2DEGs used to demonstrate USC, where the nonparabolicity of the conduction band produces a field-dependent diamagnetic coefficient, causing a redshift of the dressed cavity mode under intense THz driving. At the highest fields, the response is dominated by electron heating, but the same band curvature implies an intrinsic, coherent quartic (Kerr) term in the quantum Hopfield Hamiltonian.

\begin{figure}[t]
\centering
\includegraphics[width=0.92\columnwidth]{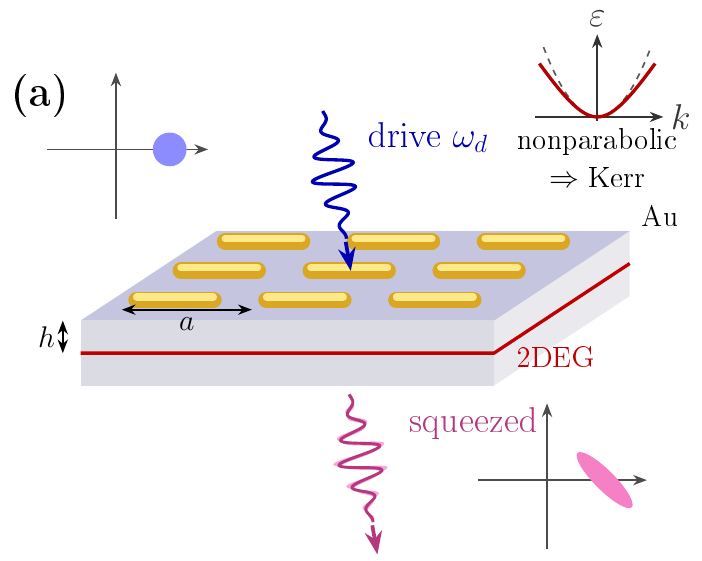}\\[2pt]
\includegraphics[width=\columnwidth]{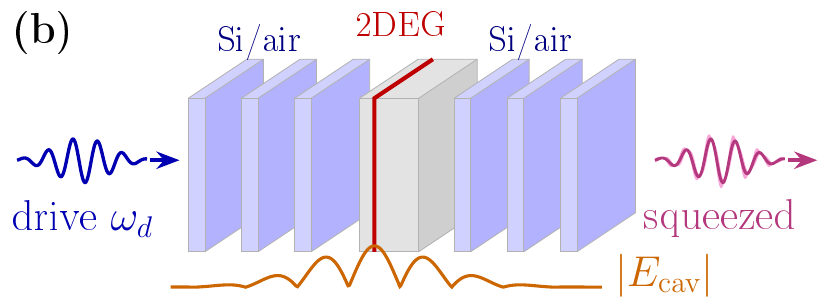}
\caption{Schematic. A narrowband THz drive at $\omega_d$ (coherent state) excites the dressed mode of a 2DEG ultrastrongly coupled to a THz cavity at $B=0$. The transmitted field is squeezed: its noise band pinches twice per cycle, and the squeezed quadrature lies at $\theta_0\simeq3\pi/4$ (phase-space insets). The Kerr nonlinearity originates from the nonparabolic conduction band (inset). (a)~Antenna-lattice metasurface a distance $h$ above the 2DEG (large $D$, moderate $Q$), a generic weakly confined geometry, distinct from the tightly confined nanoslot array of Ref.~\cite{Kim2026}; (b)~one-dimensional photonic-crystal cavity with the 2DEG (or multiple quantum wells) at the field maximum (small to moderate $D$, high $Q$, no metal loss).}
\label{fig:schematic}
\end{figure}

In this Letter we show that this intrinsic Kerr nonlinearity squeezes THz light, and that measurable squeezing is within reach of state-of-the-art samples. We develop a theory that combines (i) a quantum nonlinear Hopfield model derived from the band structure, (ii) non-rotating-wave mean-field and Gaussian-fluctuation dynamics, (iii) USC-consistent input--output with an ab initio quasinormal-mode (QNM) bath~\cite{GustinPRL2025,LSA2026}, and (iv) a self-consistent model of electron heating. 
We find that carrier heating, not the coherent physics, sets the limit; this limit is governed by a material figure of merit, which makes weakly confined cavities [Fig.~\ref{fig:schematic}(a,b)] with $Q\approx20\,\omega_{\rm cav}/D$ optimal, where $D$ is the diamagnetic coupling strength, the measure of ultrastrong coupling at zero magnetic field. We predict 1.8--2.4\,dB of output squeezing at present mobilities, and up to $\approx9$\,dB with realistic improvements.

\textit{Quantum nonlinear Hopfield model.}---We consider a 2DEG with a nonparabolic band, $\varepsilon(\mathbf p)\simeq\varepsilon_p+\alpha_{\rm np}\varepsilon_p^2$, where $\varepsilon_p=p^2/2m^*$ and $\alpha_{\rm np}=-1/E_g^*$ (two-band Kane model~\cite{Kane1957}). The electrons couple through minimal coupling, $\mathbf p\to\mathbf p+e\mathbf A$, to a single cavity mode, $\mathbf A=A_0(\aop+\ad)\mathbf e_x$, where $\mathbf e_x$ is the in-plane unit polarization vector and $A_0$ the vacuum field amplitude. At $B=0$, this yields the diamagnetic coefficient $\hbar D=Ne^2A_0^2/2m^*$, with $N$ the number of electrons in the mode, and the Hamiltonian [End Matter A]
\begin{equation}
\frac{H}{\hbar}=\omega_{\rm cav}\ad\aop+D(\aop+\ad)^2+K(\aop+\ad)^4,
\label{eq:H}
\end{equation}
with $\hbar K=\alpha_{\rm np}(\hbar D)^2/N$. At $B\neq0$, the 2DEG also contributes a cyclotron mode at $\omega_{\rm cyc}=eB/m^*$, coupled to the cavity with strength $g=g_0\sqrt{\omega_{\rm cyc}/\omega_{\rm cav}}$, where $g_0$ is the on-resonance coupling and $\eta=g_0/\omega_{\rm cav}$ the usual normalized coupling strength~\cite{Kim2026}. The Thomas--Reiche--Kuhn sum rule fixes $D=g^2/\omega_{\rm cyc}=g_0^2/\omega_{\rm cav}$, i.e., $D/\omega_{\rm cav}=\eta^2$, independent of $B$. As $B\to0$, the cyclotron frequency and the coupling $g$ vanish, but the diamagnetic term survives; thus $D/\omega_{\rm cav}$ the natural measure of USC at zero magnetic field, with $D/\omega_{\rm cav}=0.22$ ($\eta=0.47$) for the sample of Ref.~\cite{Kim2026}.
The continuous-wave (CW) drive adds $\hbar\Omega_0f(t)(\aop+\ad)$, with $f(t)=\cos\omega_dt$; similar results are obtained with multi-cycle pulses~\footnote{Gaussian pulses, $f(t)=e^{-t^2/\tau^2}\cos\omega_dt$, with $\tau\gtrsim5/\kappa$ reproduce the CW results; see the Supplemental Material.}.

The quadratic part is diagonalized by the Bogoliubov transformation $\aop=\hat c\cosh r-\hat c^\dagger\sinh r$, with $r=\tfrac12\ln R$, to a dressed mode $\hat c$ at $\omega'_{\rm cav}=R\,\omega_{\rm cav}$, $R=\sqrt{1+4D/\omega_{\rm cav}}$, whose ground state is the squeezed USC vacuum~\cite{Ciuti2005}. This system is in the USC regime even without a magnetic field, and the ground state contains $\langle\ad\aop\rangle=(R-1)^2/4R$ virtual photons, e.g., $4\times10^{-4}$, 0.025 and 0.17 for $D/\omega_{\rm cav}=0.02$, 0.22 and 1. These numbers are gauge dependent and not directly observable, since virtual photons are not emitted~\cite{Ciuti2005,DeLiberato2007}; real squeezed photons appear only under pumping, which is the subject of this Letter. For the same reason, the usual Kerr form $\chi\,\ad\ad\aop\aop$, which relies on the rotating-wave approximation (RWA) for the bare mode, cannot be used. 

The quartic term $(\aop+\ad)^4$ contains counter-rotating contributions that are not small, and must instead be expressed in the dressed basis. It acts as a Kerr nonlinearity of the dressed mode with coefficient $\chi_{\rm K}=12K/R^2$, the frequency shift per photon, plus non-RWA terms that we retain in the mean-field dynamics. For the sample of Ref.~\cite{Kim2026} ($n_{\rm 2D}=3.6\times10^{11}$\,cm$^{-2}$, $\omega_{\rm cav}/2\pi=0.87$\,THz, $\omega'_{\rm cav}/2\pi=1.19$\,THz, $D/\omega_{\rm cav}=0.22$, $E_g^*\approx2$\,eV, $N\approx1.2\times10^5$) we obtain $K/\omega_{\rm cav}\approx-7.6\times10^{-10}$. The quartic form is accurate as long as the cycle-averaged quiver energy per electron, $U_p=\langle(eA)^2\rangle/2m^*=2\hbar D\,n_c/(NR)$ (the kinetic energy of the coherent quiver motion; $\approx1$\,meV per $10^5$ photons here), is small compared with $E_g^*$; the all-orders Kane form gives identical results below $U_p\sim100$\,meV
[see the Supplemental Material (SM)~\cite{SM}]. Because the nonlinearity is shared among $N$ electrons, the shift per photon is small, $|\chi_{\rm K}|/\omega_{\rm cav}\approx5\times10^{-9}$. A coherent drive with $n_c\sim10^5$--$10^6$ photons nevertheless produces a total shift $2|\chi_{\rm K}|n_c$ comparable to the cavity linewidth.

\textit{Open-system theory and input--output.}---In the USC regime the cavity--bath coupling must be formulated in terms of dressed operators. We use the ab initio QNM result that the cavity couples to its reservoir through $\hat\Pi_c=\aop$~\cite{GustinPRL2025,LSA2026}, which gives a dressed dissipator with jump operator $\hat x^+=\hat c\cosh r$ and decay rate $\kappa=\kappa_0\cosh^2r$ [End Matter B]. The output field is $\hat b_{\rm out}=\hat b_{\rm in}-\sqrt{\kappa_0}\,\hat x^+=\hat b_{\rm in}-\sqrt{\kappa}\,\hat c$~\cite{Ridolfo2012,Salmon2022}. Applying standard input--output to the bare mode would be wrong---since $\aop=\hat c\cosh r-\hat c^\dagger\sinh r$, it predicts emission from the vacuum. The ground-state squeezing ($-1.4$\,dB in the bare quadrature) is not radiated, and all squeezing reported below refers to the dressed vacuum.

The intracavity field is macroscopic, so we write $\hat c=\alpha(t)+\delta\hat c$, solve the mean field without the rotating-wave approximation (RWA), and treat $\delta\hat c$ as Gaussian fluctuations with a linear, time-dependent covariance equation~\cite{DrummondWalls1980,CollettGardiner1984} [End Matter C]. The measurable quantity is the normalized output quadrature noise spectrum
\begin{equation}
N_\theta(\omega)=1+2\!\int\!d\tau\,e^{i\omega\tau}\,\overline{\langle\mathcal T{:}\Delta\hat X^{\rm out}_\theta(t+\tau)\Delta\hat X^{\rm out}_\theta(t){:}\rangle},
\label{eq:N}
\end{equation}
where $\mathcal T$ and ${:}\cdots{:}$ denote time and normal ordering of the output operators (appropriate for homodyne-type detection~\cite{GardinerCollett1985}), $\hat X^{\rm out}_\theta=\hat b_{\rm out}e^{-i\theta}+{\rm H.c.}$ with $\theta$ measured from the amplitude quadrature of the coherent output, and the overline is an average over the drive period; $N_\theta=1$ for vacuum. We quote $\mathcal S_{\rm out}(\omega)=10\log_{10}N_{\theta_0}(\omega)$ at the fixed angle $\theta_0$ that is optimal at the drive frequency $\omega_d$, as in a fixed-phase measurement. As a (non-measurable) reference we also use the intracavity squeezing $\mathcal S_{\rm int}=10\log_{10}[\lambda_{\min}(\mathsf V)/\tfrac12]$, where $\lambda_{\min}(\mathsf V)$ is the smallest eigenvalue of the covariance matrix of the dressed quadratures (vacuum: $\tfrac12$).

\begin{figure}[t]
\centering
\includegraphics[width=\columnwidth]{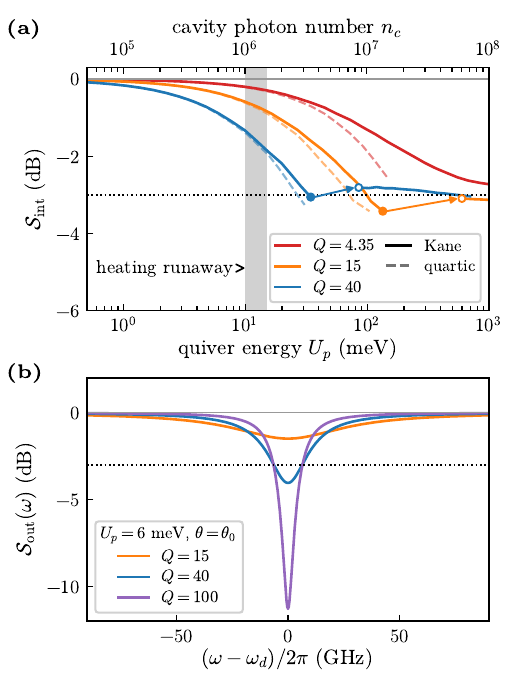}
\caption{Coherent squeezing (no heating). (a)~Intracavity squeezing $\mathcal S_{\rm int}$ of the best dressed quadrature against the cycle-averaged quiver energy $U_p$ (top axis: $n_c$) for $Q=4.35$, 15 and 40; Kane (solid) and quartic (dashed) models agree below $U_p\sim100$\,meV, and quartic curves end at their first fold or at 150\,meV, where the truncation fails [SM]. Filled circles: fold of the Kerr bistability (Kane); arrows: jump to the upper branch (open circles). $\mathcal S_{\rm int}$ peaks ($\sim$3\,dB) at the fold. Gray band: heating runaway. (b)~Output squeezing spectra $\mathcal S_{\rm out}(\omega)=10\log_{10}N_{\theta_0}(\omega)$ at the heating-compatible drive $U_p=6$\,meV ($n_c\approx6\times10^5$) for $Q=15$, 40 and 100 ($\eta=1$), reaching $-1.5$, $-4.1$ and $-11.3$\,dB at $\omega_d$, with bandwidth (FWHM) $\approx\kappa/2\pi=59$, 22 and 9\,GHz and $\Omega_0/\omega'_{\rm cav}\approx38$, 15 and 6.6. CW driving throughout, with $\omega_{\rm cav}/2\pi=0.87$\,THz, $D/\omega_{\rm cav}=0.22$ ($\omega'_{\rm cav}/2\pi=1.19$\,THz), and $\omega_d$ optimized at each point [$\omega_d/2\pi\approx1.189$\,THz in (b)].}
\label{fig:coherent}
\end{figure}

\textit{Coherent squeezing.}---Figure~\ref{fig:coherent} shows the coherent physics (CW, no heating). The intracavity squeezing grows with drive and saturates near the $\sim$3\,dB bound of a parametric process beyond the onset of bistability [Fig.~\ref{fig:coherent}(a)]. The output field, by contrast, is resolved frequency by frequency; near $\omega_d$, the field leaking from the cavity interferes destructively with the 
reflected vacuum, and the squeezing exceeds 10\,dB at $Q=100$ already at $U_p=6$\,meV, within a band of width $\sim\kappa$ [Fig.~\ref{fig:coherent}(b)]~\cite{CollettGardiner1984,CollettWalls1985}. Higher $Q$ helps because what matters is the Kerr shift relative to the linewidth, $|g|/\kappa=|\chi_{\rm K}|n_c/\kappa$. At $Q=100$, for example, the optimum requires $n_c\approx6\times10^5$ photons; the resulting shift, $2|\chi_{\rm K}|n_c/2\pi\approx5$\,GHz, is only 0.4\% of $\omega'_{\rm cav}$ but comparable to $\kappa/2\pi\approx9$\,GHz. The corresponding drive is weak, $\Omega_0/\omega'_{\rm cav}\approx8$--18 for $Q=40$--100, i.e., incident fields of only 20--30\,V/cm, two orders of magnitude below those of Ref.~\cite{Kim2026}.

\begin{figure}[t]
\centering
\includegraphics[width=\columnwidth]{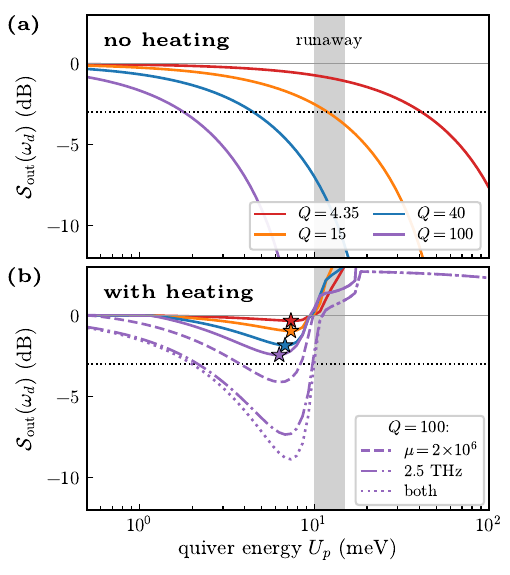}
\caption{Heating-limited squeezing. Output squeezing at $\omega_d$ against $U_p$ (CW; $\omega_d$ optimized at each point, within $\sim\kappa$ of $\omega'_{\rm cav}/2\pi=1.19$\,THz), (a)~without and (b)~with electron heating, at the sample mobility $\mu=1.2\times10^6$\,cm$^2$/Vs, for $Q=4.35$, 15, 40 and 100 ($D/\omega_{\rm cav}=0.22$, $\omega_{\rm cav}/2\pi=0.87$\,THz). Gray band: onset of runaway heating; stars: heating-limited optima. Additional curves in (b) ($Q=100$): $\mu=2\times10^6$\,cm$^2$/Vs (dashed), $\omega_{\rm cav}/2\pi=2.5$\,THz (dash-dot), and both (dotted). Dotted horizontal line: $-3$\,dB.}
\label{fig:heating}
\end{figure}

\begin{figure*}[t]
\centering
\includegraphics[width=\textwidth]{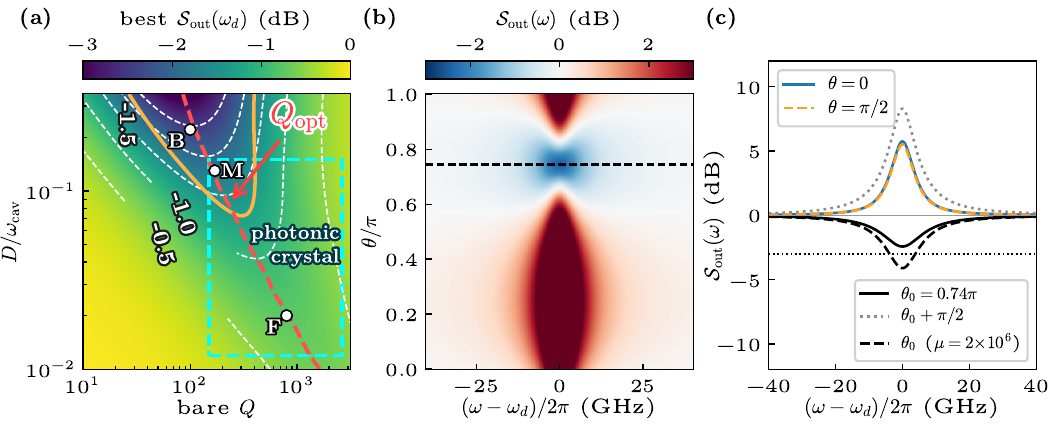}
\caption{Design rule and proposed measurement. (a)~Heating-limited best output squeezing versus bare cavity $Q$ and $D/\omega_{\rm cav}$ ($\mu=1.2\times10^6$\,cm$^2$/Vs, $\omega_{\rm cav}/2\pi=0.87$\,THz), with contours in dB. The optimum follows $Q_{\rm opt}\approx20\,\omega_{\rm cav}/D$ (red dashed). Thin orange line: $-3$\,dB contour for $\mu=2\times10^6$\,cm$^2$/Vs; inside it the squeezing exceeds 3\,dB. Markers: operating points B (sample coupling, nanoslot-like) and F (single-quantum-well photonic crystal) of Table~\ref{tab:ops}, and M, a multiple-quantum-well photonic crystal ($D/\omega_{\rm cav}\approx0.13$) at its optimal $Q$; cyan box: photonic-crystal cavities, from a single quantum well ($D/\omega_{\rm cav}\sim10^{-2}$) to multiple quantum wells ($D/\omega_{\rm cav}\approx0.13$~\cite{Li2018}). (b)~Output squeezing versus sideband frequency and quadrature angle $\theta$ (measured from the amplitude quadrature) at the optimum for $Q=100$, $D/\omega_{\rm cav}=0.22$; dashed line: $\theta_0$. (c)~Cuts at the amplitude ($\theta=0$) and phase ($\theta=\pi/2$) quadratures, at $\theta_0\approx0.74\pi$, and at $\theta_0+\pi/2$; dashed black: the squeezed quadrature for $\mu=2\times10^6$\,cm$^2$/Vs at the same coupling and $Q$ (case C of Table~\ref{tab:ops}, $-4.1$\,dB).}
\label{fig:design}
\end{figure*}

\textit{Heating limit.}---The drive inevitably heats the electrons. We describe the 2DEG by an electron temperature $T_e$, assuming fast electron--electron thermalization, with absorption through momentum scattering at rate $\gamma=e/m^*\mu$. Cooling is by LO-phonon emission from the thermal tail and, more slowly, by acoustic phonons. The electrons feed back on the cavity through $D_{\rm eff}(T_e)$, through an absorptive loss channel $\kappa_{\rm abs}=\gamma(R^2-1)/R^2$, and through its thermal noise $n_{\rm th}(T_e)$ [End Matter D]. Two regimes emerge. Below a threshold, the tiny heat capacity of the degenerate 2DEG lets $T_e$ reach a quasi-steady value of 25--40\,K within picoseconds. Once the peak quiver energy plus the thermal energy reaches $\hbar\omega_{\rm LO}=36.6$\,meV, i.e., at $U_p\approx10$--15\,meV, LO-phonon emission switches from cooling to heating and $T_e$ runs away. The squeezing must therefore be generated below this threshold [Fig.~\ref{fig:heating}].
For the mobility of the sample of Ref.~\cite{Kim2026} ($\mu=1.2\times10^6$\,cm$^2$/Vs), the best output squeezing is $-1.8$\,dB ($Q=40$) and $-2.4$\,dB ($Q=100$).
Higher mobilities, as reported in other USC samples~\cite{Scalari2012}, improve this further: $\mu=2\times10^6$\,cm$^2$/Vs gives $-4.1$\,dB. Raising the cavity frequency to 2.5\,THz gives $-7.4$\,dB, and both together give $-8.9$\,dB.

Tight field confinement is detrimental. In the 450-nm nanoslot resonator of Ref.~\cite{Kim2026}, 
the measured dressed linewidth is $\approx1.9$ times the bare one. This implies a non-radiative damping about 100 times the mobility value, consistent with nonlocal electron--hole and plasmon damping in the gap~\cite{Rajabali2021}. Including it, the heating model reproduces the redshift already observed at 0.09\,kV/cm, but leaves no squeezing [SM]. Squeezing therefore calls for weakly confined, low-loss cavities, such as photonic-crystal~\cite{Zhang2016,Li2018,Tay2025} or Fabry--Perot~\cite{Mavrona2021} cavities, or antenna lattices placed $h\gtrsim1\,\mu$m above the 2DEG, for which $qv_F/\omega\approx35\,{\rm nm}/h\ll1$.


\textit{Squeezing figure of merit and design rule.}---The origin of the limit becomes clear when the loss channels are compared at the bistability onset, where the Kerr shift is $\sim\kappa$. With $|\chi_{\rm K}|n_c=6|\alpha_{\rm np}|DU_p/R$, the ratio of absorptive to radiative loss at the largest heating-compatible drive, $U_p=U_{\max}\approx6$--8\,meV (just below the runaway onset), is [End Matter D]
\begin{equation}
\xi\equiv\frac{\kappa_{\rm abs}}{\kappa_{\rm rad}}\simeq\frac{2}{3\sqrt3}\,\frac{\gamma}{\omega_{\rm cav}}\,\frac{1}{R\,|\alpha_{\rm np}|\,U_{\max}}.
\label{eq:fom}
\end{equation}
This figure of merit is independent of the mode volume and the electron number, and of $D$ apart from $R$. It is set by the material and 
frequency---the THz scattering rate, the nonparabolicity, and the LO-phonon energy through $U_{\max}$. The collection efficiency is $\eta\simeq1/(1+\xi)$, and the thermal noise enters through the same channel. A smaller $D$ requires a proportionally smaller $\kappa$, which gives the design rule of Fig.~\ref{fig:design}(a): $Q_{\rm opt}\approx20\,\omega_{\rm cav}/D$. The best squeezing along this ridge grows with $D$, from $\approx-0.7$\,dB at $D/\omega_{\rm cav}=0.01$ to $-2.4$\,dB at 0.22 for the sample mobility of Ref.~\cite{Kim2026}. The reason is thermal: $\omega'_{\rm cav}=R\,\omega_{\rm cav}$ increases with $D$, which lowers the thermal occupation of the absorptive channel at $T_e\approx35$--40\,K, the same mechanism that favors higher cavity frequencies. Because $\xi\propto\gamma$, the whole map also improves systematically with mobility. 

At $\mu=2\times10^6$\,cm$^2$/Vs, the ridge exceeds 3\,dB for $D/\omega_{\rm cav}\gtrsim0.07$ [orange contour in Fig.~\ref{fig:design}(a)]. For specific platforms, the design rule requires $Q\sim10^3$ for photonic-crystal cavities with a single quantum well ($D/\omega_{\rm cav}\sim10^{-2}$) and $Q\sim10^2$ for metasurfaces ($D/\omega_{\rm cav}\sim0.2$). Photonic-crystal cavities with multiple quantum wells combine the advantages of both: they reach $D/\omega_{\rm cav}\approx0.13$ ($g_0/\omega_{\rm cav}=0.36$ with ten quantum wells~\cite{Li2018}) without metal loss, giving $-1.8$\,dB at the sample mobility and $-3.5$\,dB at $\mu=2\times10^6$\,cm$^2$/Vs, with $Q_{\rm opt}\approx150$ [point M in Fig.~\ref{fig:design}(a)]. These quality factors are within reach of demonstrated THz cavities~\cite{Zhang2016,Li2018,Mavrona2021}.

\textit{Potential measurement.}---The squeezed quadrature lies at $\theta_0\approx3\pi/4$ from the amplitude quadrature for all optimal operating points [Figs.~\ref{fig:design}(b) and \ref{fig:design}(c)]. Both the amplitude and the phase quadrature are anti-squeezed ($\approx+5.5$\,dB), so phase-locked detection is essential. Sub-cycle electro-optic sampling, which has resolved THz vacuum fluctuations and squeezed vacua~\cite{Riek2015,Riek2017,BeneaChelmus2019}, provides the natural detector: the field variance at a fixed sampling phase oscillates at $2\omega_d$, and two-probe correlations give $N_\theta(\omega)$. Classical precursors calibrate the prediction: Kerr hysteresis under multi-cycle driving, and phase-sensitive parametric gain of a weak probe at $\omega_d+\delta$ with an idler at $\omega_d-\delta$, which determines $|g|/\kappa$ and $\Delta/\kappa$ directly [SM].

\textit{Conclusions.}---Band nonparabolicity turns the Hopfield model of ultrastrongly coupled 2DEGs into a quantum Kerr oscillator capable of squeezing THz light, moving USC 2DEG physics beyond its classical correspondence. A USC-consistent open-system theory, together with self-consistent electron heating, predicts measurable squeezing in present samples placed in weakly confined, high-$Q$ cavities. It identifies a material figure of merit and a design rule linking $Q$ to the coupling strength. Natural extensions include Landau polaritons at $B\neq0$, where the Kane Landau ladder adds anharmonicity and quantum-Hall suppression of scattering may reduce heating; narrow-gap materials with stronger nonparabolicity; and inverse-designed cavities.

More broadly, these results take the Hopfield model, the workhorse of collective USC physics, from its classical-correspondence regime into genuine quantum optics. The linear Hopfield model is itself a quantum model, but when it is driven and probed with classical fields its observable response always has a classical counterpart. A squeezed output field does not. The same intrinsic nonlinearity opens the way to other quantum-optical effects in USC 2DEGs, such as parametric down-conversion and two-mode squeezing between the polariton branches at $B\neq0$. It also suggests the converse experiment: exciting USC polaritons with squeezed THz fields at the few-photon level, using pulse-resolved theory developed for waveguide QED~\cite{ArranzRegidor2026}, and analyzing what emerges after the interaction. A USC-based squeezed THz source would thus also serve as a probe of USC physics itself.

\begin{acknowledgments}
This work was supported by the Natural Sciences and Engineering Research Council of Canada (NSERC) (Discovery Grant and Quantum Alliance); the Canadian Foundation for Innovation (CFI); and Queen's University, Canada.
D.K. and J.K.\ acknowledge support from the U.S.\ Army Research Office (through Award Nos.\ W911NF-21-1-0157, W911NF-23-1-0410, W911NF-25-2-0150), the W.\ M.\ Keck Foundation (through Award No.\ 995764), the Gordon and Betty Moore Foundation (through Grant No.\ 11520), and the Robert A.\ Welch Foundation (through Grant No.\ C-1509). 

\end{acknowledgments}

\bibliography{prl_squeezing}

\clearpage
\section*{End Matter}

\begin{table}[b]
\caption{Model parameters. Sample values from Ref.~\cite{Kim2026}; the cavity values refer to its nanoslot resonator and are rescaled in Table~\ref{tab:ops}.}
\label{tab:params}
\begin{ruledtabular}
\footnotesize
\begin{tabular}{lll}
2DEG & $n_{\rm 2D}$, $m^*$ & $3.6\times10^{11}$\,cm$^{-2}$, $0.067\,m_e$\\
 & $E_F$ & 12.9\,meV\\
 & $\mu$ & $1.2\times10^6$\,cm$^2$/Vs\\
 & $\gamma=e/m^*\mu$ & $2.2\times10^{10}$\,s$^{-1}$\\
Band & $E_g^*$, $\alpha_{\rm np}=-1/E_g^*$ & 1.98\,eV, $-0.506$\,eV$^{-1}$\\
Cavity & $\omega_{\rm cav}/2\pi$, $D/\omega_{\rm cav}$ & 0.87\,THz, 0.22\\
 & $R$, $\omega'_{\rm cav}/2\pi$, $\cosh^2r$ & 1.372, 1.19\,THz, 1.025\\
 & $N$ (electrons in mode) & $1.17\times10^5$\\
Kerr & $K/\omega_{\rm cav}$, $\chi_{\rm K}/\omega_{\rm cav}$ & $-7.6\times10^{-10}$, $-4.8\times10^{-9}$\\
 & photons per meV of $U_p$ & $1.0\times10^5$\\
Heating & $\hbar\omega_{\rm LO}$, $\tau_{\rm LO}$ & 36.6\,meV, 0.15\,ps\\
 & $\tau_{\rm ac}$, $T_L$ & 1\,ns, 4\,K\\
 & runaway onset in $U_p$ & $\approx10$--15\,meV ($n_c\approx10^6$)\\
\end{tabular}
\end{ruledtabular}
\end{table}

\begin{table*}[t]
\caption{Heating-limited optimal CW operating points (optimized over drive and detuning). A--E: nanoslot-like coupling, $D/\omega_{\rm cav}=0.22$; F, G: photonic-crystal-like coupling, $D/\omega_{\rm cav}=0.02$. $\kappa$: dressed radiative linewidth; $\Omega_0/\omega'_{\rm cav}$: drive amplitude (incident fields $\approx20$--30\,V/cm for A--C, assuming radiation-limited input coupling); $\theta_0$: squeezed quadrature measured from the amplitude quadrature.}
\label{tab:ops}
\begin{ruledtabular}
\footnotesize
\begin{tabular}{ccccccccccccccc}
 & $\frac{D}{\omega_{\rm cav}}$ & $Q$ & $\mu$ & $\frac{\omega_{\rm cav}}{2\pi}$ & $\frac{\kappa}{2\pi}$ & $U_p$ & $n_c$ & $\frac{\Omega_0}{\omega'_{\rm cav}}$ & $T_e$ & $n_{\rm th}$ & $\eta$ & $\frac{\theta_0}{\pi}$ & $\mathcal S_{\rm int}$ & $\mathcal S_{\rm out}$\\
 & & & ($10^6$\,cm$^2$/Vs) & (THz) & (GHz) & (meV) & ($10^5$) & & (K) & & & & (dB) & (dB)\\
\hline
A & 0.22 & 40 & 1.2 & 0.87 & 22.3 & 6.8 & 6.8 & 17.8 & 39 & 0.30 & 0.87 & 0.76 & $-0.6$ & $\mathbf{-1.8}$ \\
B & 0.22 & 100 & 1.2 & 0.87 & 8.9 & 6.2 & 6.3 & 8.6 & 39 & 0.30 & 0.75 & 0.74 & $-1.0$ & $\mathbf{-2.4}$ \\
C & 0.22 & 100 & 2.0 & 0.87 & 8.9 & 6.5 & 6.5 & 8.1 & 36 & 0.25 & 0.83 & 0.75 & $-1.4$ & $\mathbf{-4.1}$ \\
D & 0.22 & 100 & 1.2 & 2.50 & 25.6 & 7.0 & 7.0 & 8.3 & 39 & 0.01 & 0.88 & 0.74 & $-2.0$ & $\mathbf{-7.4}$ \\
E & 0.22 & 100 & 2.0 & 2.50 & 25.6 & 7.2 & 7.3 & 8.1 & 36 & 0.01 & 0.92 & 0.75 & $-2.1$ & $\mathbf{-8.9}$ \\
F & 0.02 & 800 & 1.2 & 0.87 & 1.1 & 6.0 & 50.5 & 4.1 & 39 & 0.49 & 0.71 & 0.75 & $-0.4$ & $\mathbf{-0.8}$ \\
G & 0.02 & 800 & 2.0 & 0.87 & 1.1 & 6.2 & 52.6 & 3.8 & 36 & 0.43 & 0.79 & 0.75 & $-1.0$ & $\mathbf{-2.4}$ \\
\end{tabular}
\end{ruledtabular}
\end{table*}

\textit{A. Kerr Hamiltonian.}---In the two-band Kane model, $\varepsilon(1+\varepsilon/E_g^*)=p^2/2m^*$~\cite{Kane1957}, i.e., $\varepsilon\simeq\varepsilon_p+\alpha_{\rm np}\varepsilon_p^2$. With $\mathbf p\to\mathbf p+e\mathbf A$ and $\mathbf A=A_0(\aop+\ad)\mathbf e_x$, the $A^2$ part of $\varepsilon_p$ gives $\hbar D=Ne^2A_0^2/2m^*$, and $\alpha_{\rm np}(e^2A^2/2m^*)^2$ gives $\hbar K=\alpha_{\rm np}(\hbar D)^2/N$. Fermi-sea averaging renormalizes $D\to D[1+4\alpha_{\rm np}\langle\varepsilon\rangle]$, the origin of $D_{\rm eff}(T_e)$. 

Center-of-mass--relative-motion couplings (Kohn breaking~\cite{Kohn1961}) are of relative order $4|\alpha_{\rm np}|\langle\varepsilon\rangle\approx1\%$ and are neglected [SM].
 Summing all orders of the Kane dispersion gives $H_{\rm Kane}/\hbar=\omega_{\rm cav}\ad\aop+\tfrac{D^2}{2|K|}[\sqrt{1+4|K|(\aop+\ad)^2/D}-1]$, which reduces to Eq.~\eqref{eq:H} for $U_p\ll E_g^*$ [Fig.~\ref{fig:coherent}(a)] and remains bounded (monotonic) at large field, unlike the quartic truncation. Gauge consistency is ensured by applying minimal coupling to the full two-band Kane Hamiltonian, which is gauge invariant, and projecting onto the conduction band afterwards; this is equivalent to the Peierls substitution in the nonparabolic dispersion. The truncation then introduces corrections of relative order $\hbar\omega'_{\rm cav}/E_g^*\approx2.5\times10^{-3}$, avoiding the gauge ambiguities of naively truncated models~\cite{DeBernardis2018,DiStefano2019,Gustin2023} [SM].

\textit{B. Dressed input--output.}---With $\aop=\hat c\cosh r-\hat c^\dagger\sinh r$ ($e^{2r}=R$) and $\hat\Pi_c=\aop$~\cite{GustinPRL2025,LSA2026}, the dressed jump operator is $\hat x^+=\hat c\cosh r$, so $\kappa=\kappa_0\cosh^2r$, with drive strength $\Omega_0\cosh r/\sqrt R$. Field and vector-potential couplings would give $R\kappa_0$ and $\kappa_0/R$, respectively. The output, $\hat b_{\rm out}=\hat b_{\rm in}-\sqrt{\kappa_0}\,\hat x^+=\hat b_{\rm in}-\sqrt\kappa\,\hat c$, carries no flux from the ground state. We assume a one-sided (over-coupled) cavity; for a symmetric two-sided cavity observed in one port, the squeezing is diluted as $N\to1-\tfrac12(1-N)$ [SM].

\textit{C. Mean field, fluctuations and spectrum.}---With bare quadratures $\hat x=(\aop+\ad)/\sqrt2$, $\hat p=-i(\aop-\ad)/\sqrt2$, time in units of $\omega_{\rm cav}^{-1}$, and the potential $V(x)=2Dx^2+4Kx^4$ (or its Kane form), the non-RWA mean field obeys $\dot x=p+F(t)-\tfrac{\kappa}{2}x$ and $\dot p=-x-V'(x)-\tfrac{\kappa}{2}p$, with $F(t)=\sqrt2\,\Omega_0f(t)\cosh r/\sqrt R$, the dressed-consistent drive (End Matter B). Fluctuations are described by the dressed quadratures $\hat X=\sqrt R\,\delta\hat x$ and $\hat P=\delta\hat p/\sqrt R$, whose symmetrized covariance $\mathsf V$ obeys
\begin{equation}
\dot{\mathsf V}=\mathsf A\mathsf V+\mathsf V\mathsf A^{\rm T}+\mathsf N,\qquad
\mathsf A(t)=\begin{pmatrix}-\frac{\kappa}{2} & R\\[2pt] -R-\frac{V''_{\rm NL}(x(t))}{R} & -\frac{\kappa}{2}\end{pmatrix},
\label{eq:cov}
\end{equation}
with $V_{\rm NL}=V-2Dx^2$ and $\mathsf N=[\tfrac12\kappa_{\rm rad}+\kappa_{\rm abs}(n_{\rm th}+\tfrac12)]\mathsf I$. The undriven solution is $\mathsf V=\tfrac12\mathsf I$, the USC vacuum, and $\mathcal S_{\rm int}=10\log_{10}[2\lambda_{\min}(\mathsf V)]$. For the spectra, linearizing in the drive frame with $\mathbf u(\Omega)=(\delta\hat c(\Omega),\delta\hat c^\dagger(-\Omega))^{\rm T}$ gives
\begin{equation}
\mathsf M(\Omega)=\begin{pmatrix}\frac{\kappa}{2}+i(\Delta-\Omega) & ig\\ -ig^* & \frac{\kappa}{2}-i(\Delta+\Omega)\end{pmatrix},
\label{eq:M}
\end{equation}
with $\Delta=\omega'_{\rm cav}+\chi_{\rm K}(1+2n_c)-\omega_d$ and $g=\chi_{\rm K}\alpha^2$, and
\begin{equation}
N_\theta(\Omega)=\tfrac12\!\!\sum_{j={\rm rad,abs}}\!\!(2n_j+1)\sum_{k=1,2}\big|e^{-i\theta}T^{(j)}_{1k}+e^{i\theta}T^{(j)}_{2k}\big|^2,
\end{equation}
with $T^{(\rm rad)}=\mathsf I-\kappa_{\rm rad}\mathsf M^{-1}$, $T^{(\rm abs)}=-\sqrt{\kappa_{\rm rad}\kappa_{\rm abs}}\,\mathsf M^{-1}$, $n_{\rm rad}=0$ and $n_{\rm abs}=n_{\rm th}(T_e)$. This zeroth-harmonic result agrees with the non-RWA covariance to 0.1\,dB [SM].

\begin{figure}[!t]
\centering
\includegraphics[width=\columnwidth]{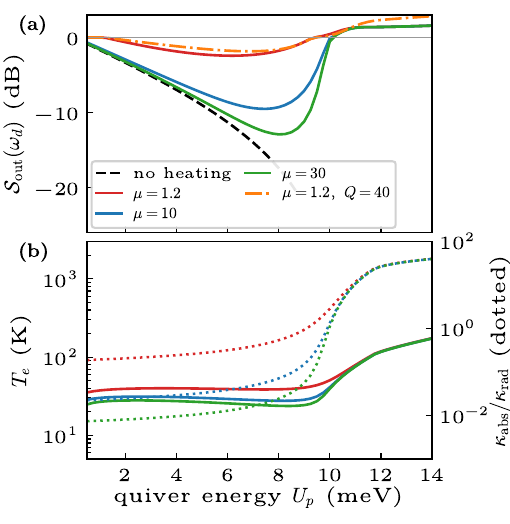}
\caption{Role of mobility ($Q=100$ unless stated, $D/\omega_{\rm cav}=0.22$). (a)~Output squeezing at $\omega_d$ against $U_p$ for several mobilities, and without heating (dashed). (b)~Electron temperature (solid, left axis) and $\kappa_{\rm abs}/\kappa_{\rm rad}$ (dotted, right axis). The temperature floor of 25--40\,K and the runaway near $U_p\approx10$\,meV are generic.}
\label{fig:EMheating}
\end{figure}

\textit{D. Heating model and figure of merit.}---The electron energy obeys
\begin{equation}
\dot U_e=2\gamma E_k+\frac{f}{\tau_{\rm LO}}(2E_k-\hbar\omega_{\rm LO})-\frac{U_e-U_e(T_L)}{\tau_{\rm ac}},
\end{equation}
with $E_k=2\hbar Dx^2/N$ and $f=\exp[-\max(\hbar\omega_{\rm LO}-\langle\varepsilon\rangle_{T_e}-E_k,0)/k_BT_e]$. The electrons feed back through $D_{\rm eff}(T_e)$, $\kappa_{\rm abs}=\gamma_{\rm eff}(R^2-1)/R^2$ with $\gamma_{\rm eff}=\gamma+f/\tau_{\rm LO}$, and $n_{\rm th}(T_e)$ [Fig.~\ref{fig:EMheating}]. We take $\gamma$ temperature independent; acoustic-phonon scattering at $T_e\approx30$--40\,K, and any difference between the THz and DC scattering rates, would increase it, an effect captured by the mobility dependence in Fig.~\ref{fig:EMheating} and Table~\ref{tab:ops}. At the bistability onset, $|\chi_{\rm K}|n_c\simeq\tfrac{\sqrt3}{2}\kappa$ together with $|\chi_{\rm K}|n_c=6|\alpha_{\rm np}|DU_p/R$ gives Eq.~\eqref{eq:fom} ($\xi\approx0.4$ for the sample), and the design rule $Q_{\rm opt}\propto\omega_{\rm cav}/D$ [Fig.~\ref{fig:design}(a)].

\textit{E. Nanoslot analysis.}---For the nanoslot sample of Ref.~\cite{Kim2026}, the dressed resonance at $B=0$ and weak field has a FWHM of 0.42\,THz, 
against 0.22\,THz at 7\,T, where the lower polariton is essentially the bare cavity mode, both extracted from the measured spectra in Fig.~1(b) of Ref.~\cite{Kim2026}. The ratio, 1.9, exceeds every bath-coupling prediction (End Matter B). Attributing $\cosh^2r\,\kappa_0$ to radiation leaves a non-radiative width of 0.20\,THz. This implies $\xi\approx0.9$ ($\eta\approx0.54$) and $\gamma_{\rm eff}=\kappa_{\rm abs}R^2/(R^2-1)\approx2.6\times10^{12}$\,s$^{-1}\approx120\gamma$; with field coupling instead, $\xi\approx0.4$. 
The implied two-port absorptance, $2\xi/(1+\xi)^2\approx0.5$, is consistent with the absorption fraction $A_0\approx0.6$ obtained from electromagnetic simulations of the nanoslot structure in Ref.~\cite{Kim2026}. A natural origin is nonlocal damping in the 450-nm gap~\cite{Rajabali2021}: for $q\simeq\pi/w$, $qv_F/\omega'_{\rm cav}\approx0.25$, and the edge fields have larger $q$. The damped energy heats the electrons. A single pulse at 0.09\,kV/cm ($U_p\approx1.5$\,meV, $\sim$0.8\,ps) deposits $\approx6$\,meV per electron ($T_e\approx100$\,K), giving a $\approx4$\,GHz redshift through $D_{\rm eff}(T_e)$ alone. This is consistent with the redshift already observed at that field, whereas with $\gamma$ alone the heating would be negligible. With this damping the heating model gives no squeezing ($\mathcal S_{\rm out}\approx+0.5$\,dB).

\textit{F. Classical precursors.}---The parameters of Eq.~\eqref{eq:M} can be measured classically before attempting squeezing. A weak probe at $\omega_d+\delta$ experiences the phase-sensitive signal gain $G_s(\delta)=|1-\kappa_{\rm rad}[\mathsf M^{-1}(\delta)]_{11}|^2$ and generates an idler at $\omega_d-\delta$ with gain $G_i(\delta)=\kappa_{\rm rad}^2|[\mathsf M^{-1}(\delta)]_{21}|^2$. Fitting both determines $|g|/\kappa$ and $\Delta/\kappa$, and hence $N_\theta(\Omega)$, without free parameters. The mean field obeys $n_c[(\kappa/2)^2+(\Delta_0+\chi_{\rm K}n_c)^2]=|\varepsilon|^2$, with $\Delta_0=\omega'_{\rm cav}+\chi_{\rm K}-\omega_d$ and $\varepsilon=\Omega_0\cosh r/(2\sqrt R)$; for $\chi_{\rm K}<0$ it becomes bistable (hysteresis under amplitude sweeps) when the drive is red detuned by $\Delta_0>\sqrt3\,\kappa/2$. Its onset calibrates $\chi_{\rm K}$, and the low-field redshift onset calibrates the heating threshold.

\begin{figure}[t]
\centering
\includegraphics[width=\columnwidth]{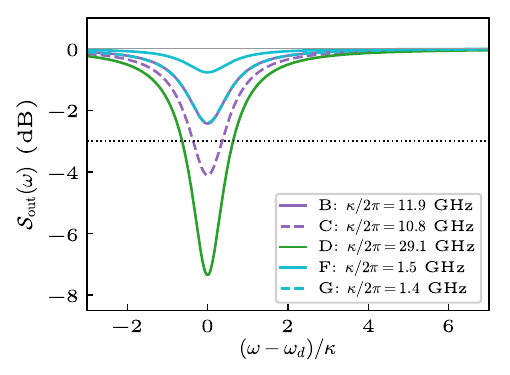}
\caption{Heated output squeezing spectra at the operating points B, C, D, F and G of Table~\ref{tab:ops}, against detuning in units of the total linewidth $\kappa=\kappa_{\rm rad}+\kappa_{\rm abs}$ (values in the legend), at fixed $\theta_0$. The photonic-crystal cases (F, G) squeeze within $\sim$1.5\,GHz, the metasurface cases (B--D) within 10--30\,GHz.}
\label{fig:EMspectra}
\end{figure}

\clearpage\clearpage   
\onecolumngrid
\setcounter{equation}{0}\setcounter{figure}{0}\setcounter{table}{0}\setcounter{section}{0}
\renewcommand{\theequation}{S\arabic{equation}}
\renewcommand{\thefigure}{S\arabic{figure}}
\renewcommand{\thetable}{S\arabic{table}}
\renewcommand{\thesection}{S\arabic{section}}
\setcounter{secnumdepth}{3}
\begin{center}
{\large\bfseries Supplementary Material for: Squeezed terahertz light from nonlinear diamagnetism in ultrastrongly coupled two-dimensional electrons}\\[6pt]
Dasom Kim, Junichiro Kono, and Stephen Hughes
\end{center}
\noindent\textit{References cited in this Supplemental Material are included in the reference list of the Letter.}

\section{Model: from the Kane band to the Kerr Hamiltonian}\label{sec:model}

\textit{Band structure.}---In the ultrastrong-coupling (USC) regime considered here, the relevant matter degree of freedom is the 2DEG in the GaAs conduction band. In the two-band Kane model its dispersion satisfies $\varepsilon(1+\varepsilon/E_g^*)=p^2/2m^*$~\cite{Kane1957}. Expanding, $\varepsilon\simeq\varepsilon_p+\alpha_{\rm np}\varepsilon_p^2$ with $\varepsilon_p=p^2/2m^*$ and $\alpha_{\rm np}=-1/E_g^*$. We use $m^*=0.067\,m_e$ and $E_g^*=1.98$\,eV, i.e., $\alpha_{\rm np}=-0.506$\,eV$^{-1}$, as in Ref.~\cite{Kim2026}.

\textit{Light--matter coupling and gauge.}---The cavity mode is described by the vector potential $\mathbf A=A_0(\aop+\ad)\mathbf e_x$, uniform over the 2DEG, where $\mathbf e_x$ is the in-plane unit polarization vector and $A_0$ the vacuum field amplitude. Minimal coupling $\mathbf p\to\mathbf p+e\mathbf A$ is applied to the full two-band Kane Hamiltonian, which is gauge invariant, before projecting onto the conduction band. This is equivalent to the Peierls substitution $\varepsilon(\mathbf p)\to\varepsilon(\mathbf p+e\mathbf A)$ in the nonparabolic dispersion. Projecting first and coupling afterwards would break gauge invariance~\cite{DeBernardis2018,DiStefano2019,Gustin2023}. The neglected interband terms are of relative order $\hbar\omega'_{\rm cav}/E_g^*\approx2.5\times10^{-3}$.

\textit{Fermi-sea average.}---Summing $\varepsilon(\mathbf p_j+e\mathbf A)$ over the $N$ electrons in the mode and expanding in $A$ gives (i) a term linear in $A$ that vanishes for the equilibrium Fermi sea; (ii) the diamagnetic term $\hbar D(\aop+\ad)^2$, with
\begin{equation}
\hbar D=\frac{Ne^2A_0^2}{2m^*}\,\big[1+4\alpha_{\rm np}\langle\varepsilon\rangle\big],
\end{equation}
where the bracket is the occupation-averaged curvature (absorbed into the measured $D$ at $T_L$, and the source of $D_{\rm eff}(T_e)$ under heating); and (iii) the quartic term $\hbar K(\aop+\ad)^4$ with $\hbar K=\alpha_{\rm np}(\hbar D)^2/N$. 
Terms that couple the center-of-mass motion to relative (intraband) motion, and so break Kohn's theorem~\cite{Kohn1961}, are of relative order $4|\alpha_{\rm np}|\langle\varepsilon\rangle$, the same order as the renormalization $D_{\rm eff}-D_{\rm par}$. For the degenerate 2DEG, $\langle\varepsilon\rangle\approx E_F/2\approx6.5$\,meV (about 7\,meV at $T_e\approx40$\,K), giving $\approx1.3$--1.4\%, and these terms are neglected.

Using instead the full Kane dispersion, $\varepsilon(p)=\tfrac{E_g^*}{2}\big[\sqrt{1+4\varepsilon_p/E_g^*}-1\big]$ (the solution of the Kane relation above, with $-1$ setting $\varepsilon(0)=0$), and summing over the Fermi sea to all orders in $A$ gives
\begin{equation}
\frac{H_{\rm Kane}}{\hbar}=\omega_{\rm cav}\ad\aop+\frac{D^2}{2|K|}\left[\sqrt{1+\frac{4|K|}{D}(\aop+\ad)^2}-1\right],
\label{eq:SKane}
\end{equation}
where the $-1$ removes the field-independent constant, so that the expansion in $u=(\aop+\ad)^2$, $\tfrac{D^2}{2|K|}[\sqrt{1+4|K|u/D}-1]=Du-|K|u^2+\dots$, starts with the diamagnetic term and reproduces Eq.~(1) of the main text ($K<0$). Unlike the quartic truncation, Eq.~\eqref{eq:SKane} remains monotonic in the field (no spurious barrier).

\begin{figure}[h]
\centering\includegraphics[width=0.8\textwidth]{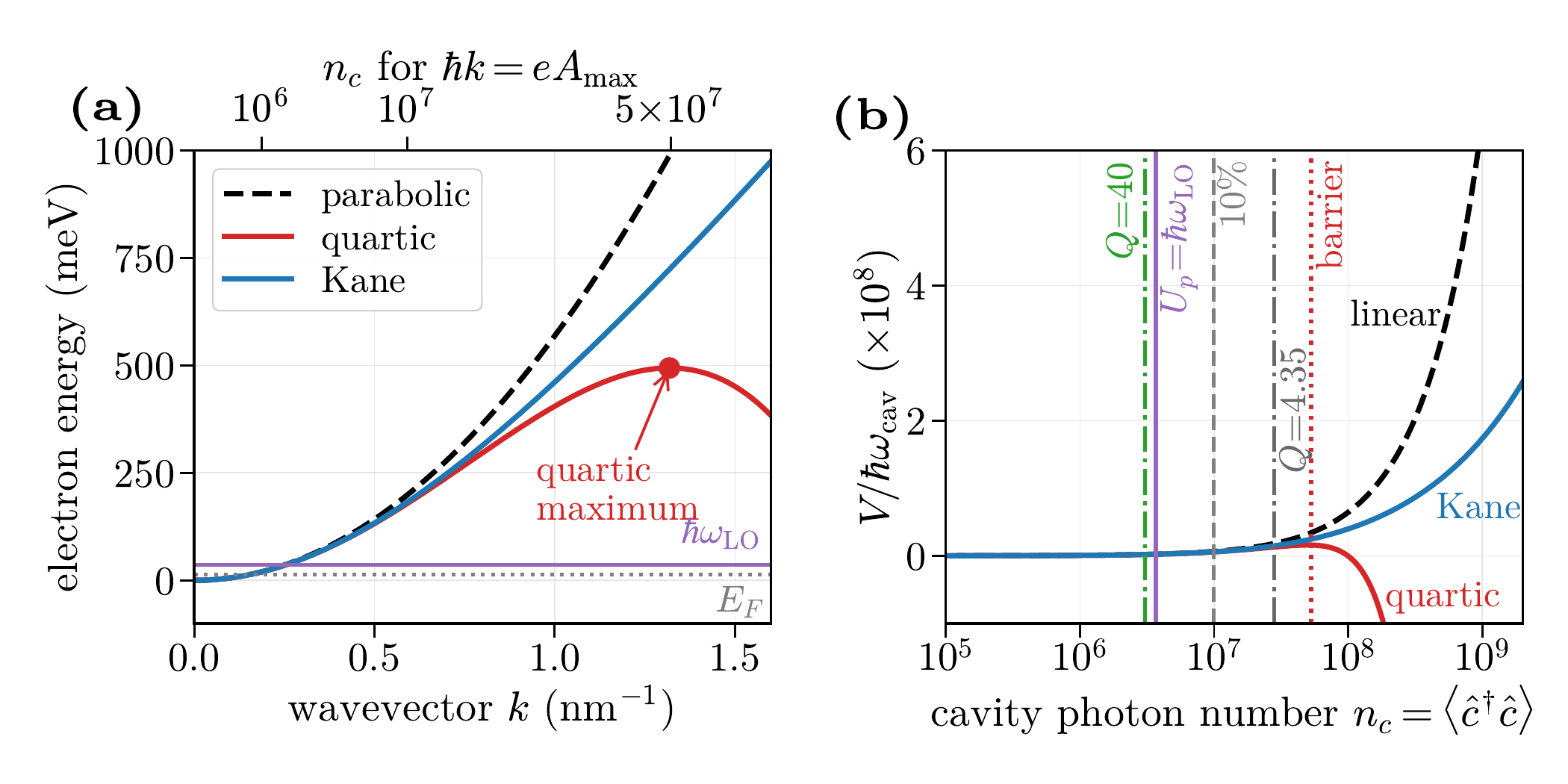}
\caption{Band nonparabolicity and cavity potential. (a)~Kane dispersion compared with the parabolic band and the quartic truncation, with the LO-phonon energy $\hbar\omega_{\rm LO}=36.6$\,meV (solid) and the Fermi energy $E_F=12.9$\,meV at $T=0$ (dotted); the top axis gives the cavity photon number $n_c$ at which the quiver momentum $eA_{\max}$ equals $\hbar k$. (b)~The cavity potential $V(x)$ for the quartic truncation (which develops a barrier at large field) and for the all-orders Kane form of Eq.~\eqref{eq:SKane} (monotonic).}
\label{fig:Sband}
\end{figure}

\textit{Parameters and scales.}---For the sample of Ref.~\cite{Kim2026} ($n_{\rm 2D}=3.6\times10^{11}$\,cm$^{-2}$, $\omega_{\rm cav}/2\pi=0.87$\,THz, $D/\omega_{\rm cav}=0.22$, $N=1.17\times10^5$) we find $K/\omega_{\rm cav}=-7.6\times10^{-10}$ and $\chi_{\rm K}/\omega_{\rm cav}=-4.8\times10^{-9}$ (Table~I in the End Matter of the main text). The characteristic drive scales are listed in Table~\ref{tab:scales}. Figure~\ref{fig:Sband} compares the quartic and Kane potentials.

\begin{table}[h]
\caption{Characteristic scales in terms of the cycle-averaged quiver energy $U_p=2\hbar Dn/(NR)$ and the dressed photon number $n_c$ (sample parameters).}
\label{tab:scales}
\begin{ruledtabular}
\begin{tabular}{lll}
Scale & Criterion & Value\\ \hline
Onset of runaway heating & $2U_p+\langle\varepsilon\rangle\simeq\hbar\omega_{\rm LO}$ & $U_p\approx10$--15\,meV, $n_c\approx10^6$\\
Heating-limited optimum & Table~II of the main text & $U_p\approx6$--8\,meV, $n_c\approx(6$--$8)\times10^5$\\
Crude LO criterion & $U_p=\hbar\omega_{\rm LO}=36.6$\,meV & $n_c\approx3.7\times10^6$\\
Bistability onset & $|\chi_{\rm K}|n_c\simeq\tfrac{\sqrt3}{2}\kappa$ & $n_c\approx2.8\times10^7$ ($Q=4.35$), $3.0\times10^6$ ($Q=40$)\\
10\% quartic correction to $D$ & & $n_c\approx10^7$\\
Quartic barrier (truncation fails) & $V'(x)=0$ & $n_c\approx5.3\times10^7$\\
\end{tabular}
\end{ruledtabular}
\end{table}

\section{Dressed operators, cavity bath and input--output}\label{sec:bath}

\textit{Dressed mode.}---The quadratic part of the Hamiltonian [Eq.~(1) of the main text, or $H_{\rm Kane}$ of Eq.~\eqref{eq:SKane}], $\hbar\omega_{\rm cav}\ad\aop+\hbar D(\aop+\ad)^2$, is diagonalized by the Bogoliubov transformation $\aop=\hat c\cosh r-\hat c^\dagger\sinh r$ with $e^{2r}=R=\sqrt{1+4D/\omega_{\rm cav}}$. Since $\aop+\ad=(\hat c+\hat c^\dagger)/\sqrt R$, the full Hamiltonian becomes
\begin{equation}
\frac{H}{\hbar}=\omega'_{\rm cav}\hat c^\dagger\hat c+\frac{K}{R^2}(\hat c+\hat c^\dagger)^4
\simeq(\omega'_{\rm cav}+\chi_{\rm K})\hat c^\dagger\hat c+\frac{\chi_{\rm K}}{2}\hat c^\dagger\hat c^\dagger\hat c\hat c+\big[\text{non-RWA terms}\big],
\label{eq:Sdressed}
\end{equation}
up to a constant, with $\omega'_{\rm cav}=R\omega_{\rm cav}$ and $\chi_{\rm K}=12K/R^2$. The non-RWA terms ($\hat c^4$, $\hat c^{\dagger3}\hat c$, and similar) are retained in the mean-field dynamics. Equivalently, with $\hat x=(\aop+\ad)/\sqrt2$ and $\hat p=-i(\aop-\ad)/\sqrt2$, the dressed quadratures are $\hat X=\sqrt R\,\hat x$ and $\hat P=\hat p/\sqrt R$.

\textit{Real and virtual photons.}---Throughout, $n_c=\langle\hat c^\dagger\hat c\rangle$ denotes the number of real (dressed) cavity photons, essentially the coherent amplitude $|\alpha|^2$ generated by the drive; it vanishes in the ground state. This is distinct from the virtual photon number of the USC ground state, $\langle\ad\aop\rangle_{\rm GS}=\sinh^2r=(R-1)^2/4R$ (0.025 for $R=1.372$), which corresponds to $-1.37$\,dB of squeezing in the bare $\hat x$ quadrature. 
%
Virtual photons are gauge dependent~\cite{DeBernardis2018,DiStefano2019},  so the ground-state photon number depends on which field operator is identified as ``the photon'' and is therefore not an observable. Only gauge-invariant quantities, such as the field fluctuations or the emitted photon flux, are physically meaningful, and virtual photons are not emitted. Only real photons, created by pumping, reach the detector.

\textit{Cavity--bath coupling.}---The cavity mode is a lossy quasinormal mode (QNM), i.e., an eigenmode of the open electromagnetic problem with complex frequency. From the quantized QNM theory of USC~\cite{GustinPRL2025,LSA2026}, the cavity couples to its radiative reservoir through the operator $\hat\Pi_c=\aop$, with the system--bath Hamiltonian
\begin{equation}
H_{\rm SB}=\hbar\int d\omega\,\sqrt{\frac{\kappa_0}{2\pi}}\,\big[\hat b^\dagger(\omega)\hat\Pi_c+{\rm H.c.}\big],
\end{equation}
where $\hat b(\omega)$ annihilates a reservoir (output-channel) photon at frequency $\omega$, with $[\hat b(\omega),\hat b^\dagger(\omega')]=\delta(\omega-\omega')$ and $[\hat b(\omega),\hat b(\omega')]=0$, and $\kappa_0$ is the bare-cavity decay rate. The operator $\hat\Pi_c$ need not be Hermitian; only $H_{\rm SB}$ must be. In the dressed basis, keeping only the system transitions that the reservoir can absorb (positive dressed frequencies), the jump operator is $\hat x^+=\hat c\cosh r$, the dressed decay rate is $\kappa=\kappa_0\cosh^2r=1.025\,\kappa_0$, and a classical drive coupled through the same port acts with strength $\Omega_0\cosh r/\sqrt R=0.864\,\Omega_0$. Table~\ref{tab:bath} compares this with other choices of coupling operator. The ratio of dressed to bare linewidth therefore tests the cavity--bath coupling, provided non-radiative damping is negligible; Sec.~\ref{sec:nanoslot} shows that it is not negligible in the nanoslot.

\begin{table}[h]
\caption{Dressed decay rate for different cavity--bath coupling operators ($R=1.372$).}
\label{tab:bath}
\begin{ruledtabular}
\begin{tabular}{lll}
Coupling operator & $\kappa/\kappa_0$ & Value\\ \hline
$\hat\Pi_c=\aop$ (ab initio QNM) & $\cosh^2r$ & 1.025\\
electric-field coupling & $R$ & 1.37\\
vector-potential coupling & $1/R$ & 0.73\\
RWA (circuit-QED style) & $\cosh^2r$ & 1.025\\
\end{tabular}
\end{ruledtabular}
\end{table}

\textit{Input--output.}---For a single radiative channel the output field is
\begin{equation}
\hat b_{\rm out}(t)=\hat b_{\rm in}(t)-\sqrt{\kappa_0}\,\hat x^+(t)=\hat b_{\rm in}(t)-\sqrt{\kappa}\,\hat c(t),
\label{eq:Sio}
\end{equation}
where $\hat b_{\rm in,out}$ are defined from $\hat b(\omega)$ before and after the interaction~\cite{GardinerCollett1985}, $\hat x^+=\hat c\cosh r$ is the de-excitation (positive-frequency) part of $\hat\Pi_c$ in the dressed basis, and $\kappa=\kappa_0\cosh^2r$. Two common shortcuts fail in the USC regime.
\begin{enumerate}
\item The bare-operator relation, $\hat b_{\rm out}=\hat b_{\rm in}-\sqrt{\kappa_0}\,\aop$, contains $-\sqrt{\kappa_0}\,\hat c^\dagger\sinh r$. This predicts a photon flux $\kappa_0\sinh^2r$ out of the ground state, violating energy conservation; the counter-rotating term cannot be resonant with any positive-frequency reservoir mode and contributes only virtual corrections.
\item The widely used dressed prescription takes $\hat x^+$ as the positive-frequency part of a cavity \emph{field} quadrature~\cite{Ridolfo2012}. This removes the ground-state flux, but the result depends on which quadrature is assumed to couple to the reservoir: the electric field or the vector potential give different decay rates and drive factors (Table~\ref{tab:bath}).
\end{enumerate}
The quantized QNM theory resolves this ambiguity by fixing $\hat\Pi_c=\aop$~\cite{GustinPRL2025,LSA2026}. USC input--output, including these differences and their experimental signatures, will be discussed in 
future work.

\textit{Ports, symmetric and one-sided cavities.}---With several radiative channels $j$ (e.g., the two mirrors of a Fabry--Perot or photonic-crystal cavity), Eq.~\eqref{eq:Sio} generalizes to $\hat b_{{\rm out},j}=\hat b_{{\rm in},j}-\sqrt{\kappa_{0,j}}\,\hat x^+=\hat b_{{\rm in},j}-\sqrt{\kappa_j}\,\hat c$, with $\kappa_j=\kappa_{0,j}\cosh^2r$ and $\kappa_{\rm rad}=\sum_j\kappa_j$. The vacuum entering the unobserved ports dilutes the squeezing seen in port $j$, $N_j=1-\eta_j(1-N)$, with port efficiency $\eta_j=\kappa_j/\kappa$. For a symmetric two-sided cavity measured in transmission, $\eta_j=1/2$. The results in the main text assume a one-sided (or strongly over-coupled) cavity, in which the radiated field leaves through a single collected port ($\eta_j\simeq\kappa_{\rm rad}/\kappa$). This corresponds to a photonic-crystal cavity with one mirror much more reflective than the other, a cavity measured in reflection with a back reflector, or a chiral emitter that radiates into a single direction. With an absorptive channel (electrons, metal) of rate $\kappa_{\rm abs}$, the absorptive port acts as an additional, thermal input (see End Matter C of the main text).

\section{Mean field, Gaussian fluctuations and validation}\label{sec:method}

\textit{Equations.}---With time in units of $\omega_{\rm cav}^{-1}$, bare quadratures $\hat x=(\aop+\ad)/\sqrt2$ and $\hat p=-i(\aop-\ad)/\sqrt2$, and $V(x)=2Dx^2+4Kx^4$ [or the Kane form, Eq.~\eqref{eq:SKane}, with $(\aop+\ad)^2\to2x^2$], the non-RWA mean field and the dressed covariance obey
\begin{align}
\dot x&=p+F(t)-\tfrac{\kappa}{2}x,\qquad \dot p=-x-V'(x)-\tfrac{\kappa}{2}p,\label{eq:Smf}\\
\dot{\mathsf V}&=\mathsf A\mathsf V+\mathsf V\mathsf A^{\rm T}+\mathsf N,\qquad
\mathsf A(t)=\begin{pmatrix}-\kappa/2 & R\\ -R-V''_{\rm NL}(x(t))/R & -\kappa/2\end{pmatrix},\label{eq:Scov}
\end{align}
with $F(t)=\sqrt2\,\Omega_0f(t)\cosh r/\sqrt R$, $V_{\rm NL}=V-2Dx^2$, and the noise matrix $\mathsf N$ given in Eq.~\eqref{eq:Snoise} below.

\textit{Derivation of the covariance equation.}---With $\hbar=\omega_{\rm cav}=1$, the Hamiltonian of Eq.~(1) of the main text in bare quadratures reads $H=\tfrac12(\hat x^2+\hat p^2)+V(\hat x)$. Using $1+4D=R^2$ and the dressed quadratures $\hat X=\sqrt R\,\hat x$ and $\hat P=\hat p/\sqrt R$, this becomes
\begin{equation}
H=\frac R2\big(\hat X^2+\hat P^2\big)+V_{\rm NL}\big(\hat X/\sqrt R\big),
\label{eq:SHdressed}
\end{equation}
i.e., a dressed oscillator at $\omega'_{\rm cav}=R$ plus the nonlinear part $V_{\rm NL}$. Dissipation acts on the dressed mode $\hat c=(\hat X+i\hat P)/\sqrt2$ (Sec.~\ref{sec:bath}). The radiative channel at rate $\kappa_{\rm rad}$ is at zero temperature, and the absorptive (electronic) channel at rate $\kappa_{\rm abs}$ has thermal occupation $n_{\rm th}$. The quantum Langevin equations are then~\cite{GardinerCollett1985,CollettGardiner1984}
\begin{equation}
\dot{\hat X}=R\hat P-\frac{\kappa}{2}\hat X+\hat\xi_X,\qquad
\dot{\hat P}=-R\hat X-\frac{1}{\sqrt R}V_{\rm NL}'\big(\hat X/\sqrt R\big)-\frac{\kappa}{2}\hat P+\hat\xi_P,
\label{eq:SLangevin}
\end{equation}
with $\kappa=\kappa_{\rm rad}+\kappa_{\rm abs}$. The noise operators $\hat\xi_{X,P}$ have zero mean, and their symmetrized correlations are
\begin{equation}
\tfrac12\big\langle\{\hat\xi_i(t),\hat\xi_j(t')\}\big\rangle=N_{ij}\,\delta(t-t'),\qquad \mathsf N=\Big[\tfrac12\kappa_{\rm rad}+\kappa_{\rm abs}\big(n_{\rm th}+\tfrac12\big)\Big]\mathsf I,
\label{eq:Snoise}
\end{equation}
where each channel contributes its rate times its symmetrized occupation, $n+\frac12$. The coherent drive enters through the radiative port as a classical force on the quadratures, which gives the term $F(t)$ in Eq.~\eqref{eq:Smf}; for CW driving its phase relative to the quadratures only shifts the time origin.

We now write $\hat X=\bar X(t)+\delta\hat X$ and $\hat P=\bar P(t)+\delta\hat P$, where the mean fields obey the classical equations, i.e., Eq.~\eqref{eq:Smf} in bare quadratures, and expand Eq.~\eqref{eq:SLangevin} to first order in the fluctuations~\cite{DrummondWalls1980,CollettGardiner1984}. With $V_{\rm NL}'(\hat x)\simeq V_{\rm NL}'(\bar x)+V_{\rm NL}''(\bar x)\,\delta\hat X/\sqrt R$, the fluctuations obey the linear equation
\begin{equation}
\delta\dot{\hat{\mathbf R}}=\mathsf A(t)\,\delta\hat{\mathbf R}+\hat{\boldsymbol\xi},\qquad
\delta\hat{\mathbf R}=\begin{pmatrix}\delta\hat X\\ \delta\hat P\end{pmatrix},\qquad
\hat{\boldsymbol\xi}=\begin{pmatrix}\hat\xi_X\\ \hat\xi_P\end{pmatrix},
\label{eq:Slinear}
\end{equation}
with $\mathsf A(t)$ as in Eq.~\eqref{eq:Scov}. The curvature of the nonlinear potential, evaluated on the classical trajectory, thus acts as a time-periodic modulation of the oscillator's restoring force: an internally generated parametric pump at $2\omega_d$, which squeezes the fluctuations. For the covariance matrix $V_{ij}=\tfrac12\langle\{\delta\hat R_i,\delta\hat R_j\}\rangle$, Eq.~\eqref{eq:Slinear} gives $\dot{\mathsf V}=\langle\delta\dot{\hat{\mathbf R}}\,\delta\hat{\mathbf R}^{\rm T}\rangle_{\rm sym}+\langle\delta\hat{\mathbf R}\,\delta\dot{\hat{\mathbf R}}^{\rm T}\rangle_{\rm sym}$. The drift terms give $\mathsf A\mathsf V+\mathsf V\mathsf A^{\rm T}$ and, because the noise is delta correlated, the noise terms give $\mathsf N$, which is Eq.~\eqref{eq:Scov}. Without drive, $\mathsf A=-\tfrac\kappa2\mathsf I+R\,\mathsf J$ (with $\mathsf J$ the symplectic matrix), and the steady state is $\mathsf V=\tfrac12\mathsf I$ for $n_{\rm th}=0$, the dressed vacuum. For a linear system with Gaussian noise, the state remains Gaussian and is fully determined by $(\bar X,\bar P)$ and $\mathsf V$. The intracavity squeezing of the best quadrature is $\mathcal S_{\rm int}=10\log_{10}(2\lambda_{\min})$, with $\lambda_{\min}$ the smaller eigenvalue of $\mathsf V$. The neglected higher-order terms in $\delta\hat X$ produce non-Gaussian corrections of relative order $|\chi_{\rm K}|/\kappa$; see the last paragraph of this section.

We integrate Eqs.~\eqref{eq:Smf}--\eqref{eq:Scov} with fourth-order Runge--Kutta at 70 steps per drive period. CW results are taken after $\geq12$ cavity decay times, and the drive frequency is optimized on a grid finer than $0.06\kappa$.

\begin{figure}[h]
\centering\includegraphics[width=0.85\textwidth]{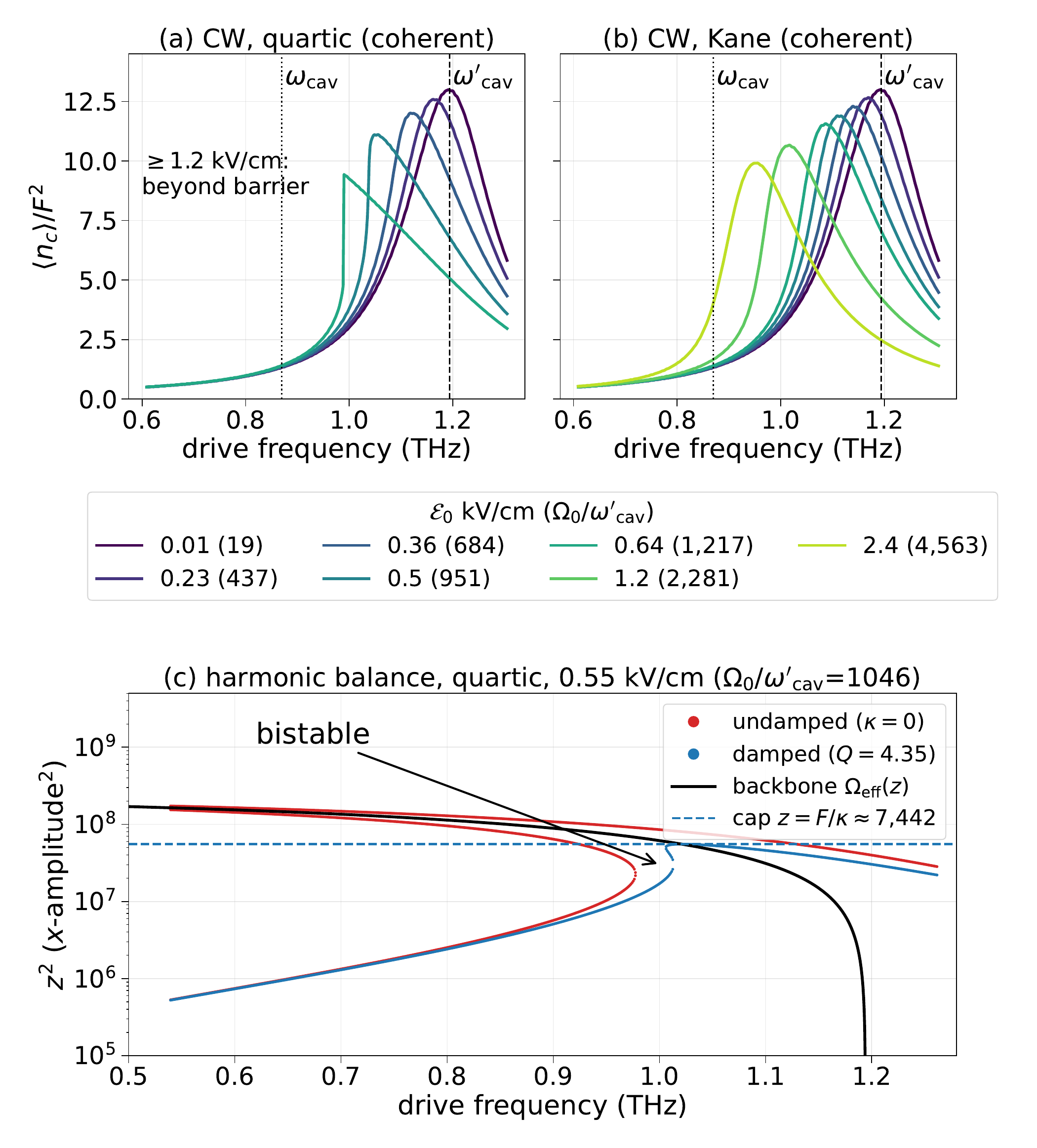}
\caption{CW mean-field response for $Q=4.35$. (a,b) Normalized intracavity photon number against drive frequency for the quartic and Kane models at several incident fields (legend: $\mathcal E_0$ and $\Omega_0/\omega'_{\rm cav}$). (c) Harmonic balance of the quartic model above the fold (0.55\,kV/cm): undamped, damped, backbone, and the damping cap $z=F/\kappa$; the damped response is S-shaped (bistable).}
\label{fig:Scw}
\end{figure}

\textit{Mean-field response.}---Figure~\ref{fig:Scw} shows the CW response. The quartic model shows the softening and asymmetric resonance of a Duffing oscillator~\cite{LandauMechanics,NayfehMook1979} and, above $\approx0.5$\,kV/cm for $Q=4.35$, bistability. The Kane response softens monotonically, and its total shift is bounded because $D_{\rm eff}\to0$. Damping is essential, since without it, harmonic balance gives a backbone with no well-defined resonance. Pulses produce shifts about six times smaller than CW driving at the same peak field.

\begin{figure}[h]
\centering\includegraphics[width=0.9\textwidth]{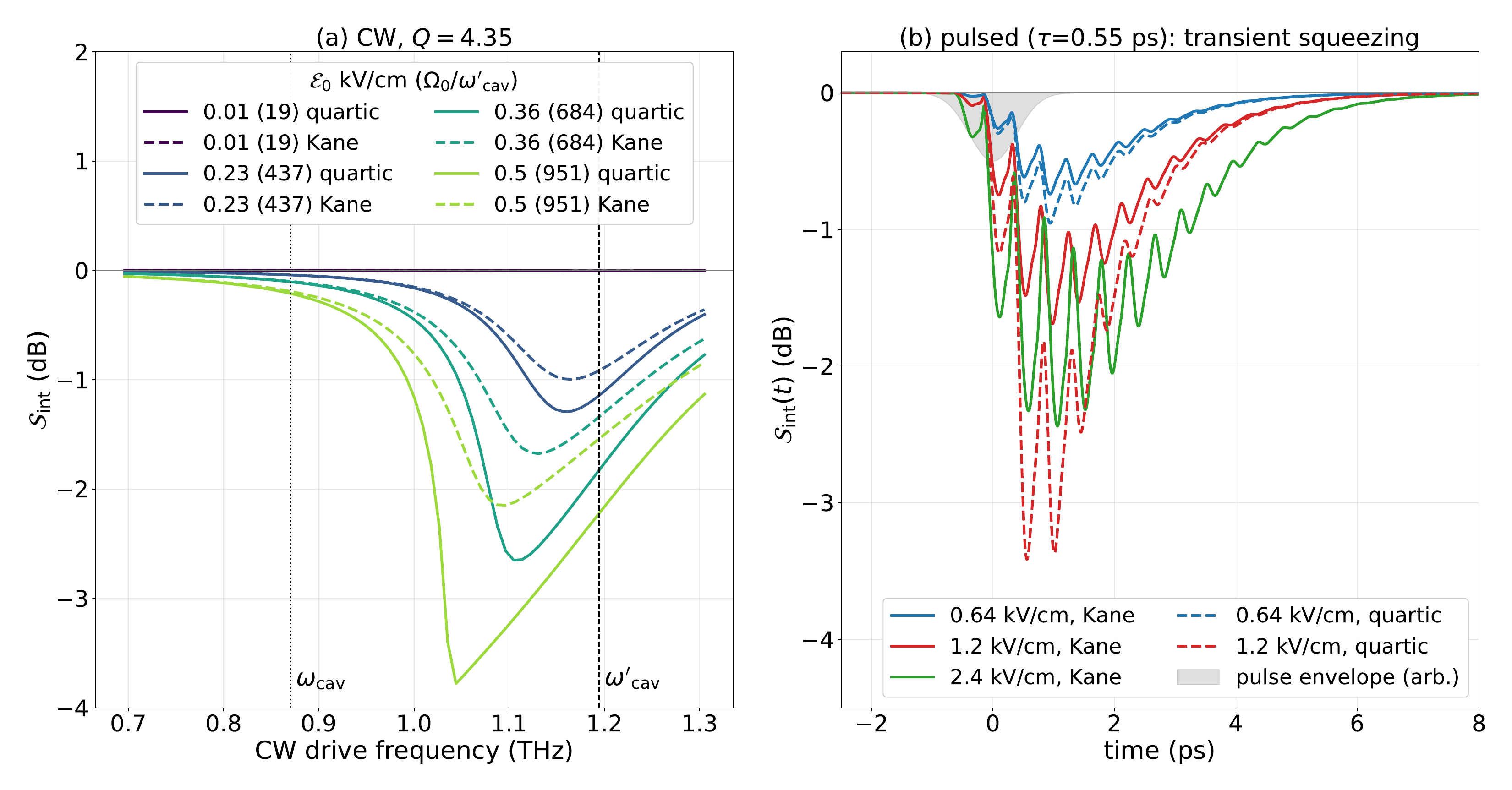}
\caption{Intracavity squeezing of the best dressed quadrature (coherent dynamics, $Q=4.35$). (a)~CW, against drive frequency, for the quartic and Kane models. (b)~Gaussian pulses, $f(t)=e^{-t^2/\tau^2}\cos\omega_dt$ with $\tau=0.55$\,ps (field-envelope FWHM $\approx0.92$\,ps), against time; gray: pulse envelope.}
\label{fig:Ssq}
\end{figure}

\textit{Intracavity squeezing, CW and pulsed.}---Figure~\ref{fig:Ssq} shows $\mathcal S_{\rm int}$ for CW and Gaussian pulses ($f(t)=e^{-t^2/\tau^2}\cos\omega_dt$). For single-cycle pulses with $\tau=0.55$\,ps, i.e., a field-envelope FWHM $\tau_{\rm FWHM}=2\sqrt{\ln2}\,\tau\approx0.92$\,ps (intensity FWHM $\sqrt{2\ln2}\,\tau\approx0.65$\,ps), at the fields of Ref.~\cite{Kim2026}, the coherent dynamics alone would give a transient $-0.7$, $-1.7$ and $-2.4$\,dB at 0.64, 1.2 and 2.4\,kV/cm. These fields lie far above the runaway threshold, however, so this squeezing is not accessible in practice. Multi-cycle pulses with $\tau\gtrsim5/\kappa$ reproduce the CW results.

\textit{Validation of the drive-frame (zeroth-harmonic) approximation.}---The output spectra and heating-limited optima are computed by linearizing the fluctuations in the frame of the drive (see End Matter C of the main text), which keeps only the zeroth Floquet harmonic. At the operating points the $2\omega_d$ modulation of $V''_{\rm NL}$ is only a few per cent. Table~\ref{tab:valid} compares the resulting intracavity squeezing with the full non-RWA covariance of Eq.~\eqref{eq:Scov}; the two agree to $\approx0.1$\,dB.

\begin{table}[h]
\caption{Intracavity squeezing from the drive-frame model and from the non-RWA covariance equation (coherent dynamics, $U_p=\hbar\omega_{\rm LO}$ reference points).}
\label{tab:valid}
\begin{ruledtabular}
\begin{tabular}{ccc}
$Q$ & drive frame (dB) & non-RWA (dB)\\ \hline
4.35 & $-0.60$ & $-0.64$\\
15 & $-1.71$ & $-1.66$\\
40 & $-2.72$ & $-2.81$\\
\end{tabular}
\end{ruledtabular}
\end{table}

\textit{Non-RWA output spectra.}---Floquet theory is not required to compute the output spectrum. For the linear, time-periodic fluctuation equation~\eqref{eq:Slinear} with delta-correlated noise, the two-time correlations follow from the quantum regression theorem,
\begin{equation}
\langle\delta\hat{\mathbf R}(t+\tau)\,\delta\hat{\mathbf R}^{\rm T}(t)\rangle=\Phi(t+\tau,t)\,\langle\delta\hat{\mathbf R}(t)\,\delta\hat{\mathbf R}^{\rm T}(t)\rangle,\qquad \partial_\tau\Phi(t+\tau,t)=\mathsf A(t+\tau)\,\Phi(t+\tau,t),
\end{equation}
for $\tau\geq0$, with $\Phi(t,t)=\mathsf I$. Averaging over $t$ within a drive period, applying the input--output relation, and Fourier transforming in $\tau$ gives $N_\theta(\omega)$ of Eq.~(2) of the main text without a rotating-wave approximation or a Floquet truncation. The result is exact within the Gaussian theory and the Born--Markov treatment of the reservoirs, which underlies both the delta-correlated noise of Eq.~\eqref{eq:Snoise} and the regression theorem. The same procedure has been used for squeezed pulses~\cite{ArranzRegidor2026}. The Markov approximation fails for a structured (non-Lorentzian) environment whose spectral density varies over the bandwidth probed by the dynamics, for example under very short pulses or strong modulation, where the Floquet sidebands sample the bath at different frequencies. The system--bath rates must then be evaluated at the Floquet quasienergy transitions, which requires a Floquet treatment of the dissipator~\cite{Akbari2025,AkbariFloquet2026}. For the nearly flat spectral densities of the low-$Q$ cavities and the Drude damping considered here, and for CW or multi-cycle driving, the Markov approximation is well justified.

\textit{Gaussian approximation.}---Non-Gaussian corrections scale with the Kerr shift per photon relative to the linewidth, $|\chi_{\rm K}|/\kappa\sim10^{-6}$ for $Q=100$, and are negligible. This can be checked directly by displacing $\aop=\alpha(t)+\delta\aop$ and solving the master equation for $\delta\aop$ in a small Fock space, where the Kerr term yields cubic and quartic terms in $\delta\aop$.

\section{Electron heating}\label{sec:heating}

\textit{Thermodynamics.}---Electron--electron scattering thermalizes the 2DEG in well under a picosecond at this density, so we describe it by a temperature $T_e$, using 2D Fermi--Dirac statistics at fixed density ($E_F=12.9$\,meV). The mean kinetic energy $U(T_e)$ and its inverse are tabulated numerically. The electronic heat capacity of the degenerate gas is very small, $C_e=(\pi^2/3)k_B^2T_e/E_F\approx7\times10^{-3}$\,meV/K per electron at 4\,K.

\textit{Energy balance.}---The coherent quiver motion has instantaneous kinetic energy $E_k=(eA)^2/2m^*=2\hbar Dx^2/N$ per electron, whose cycle average is $U_p$. The energy balance, Eq.~(7) in End Matter D of the main text, contains three processes:
\begin{itemize}
\item \emph{Momentum scattering}, at rate $\gamma=e/m^*\mu$, randomizes the quiver momentum and absorbs $2\gamma E_k$ (Drude absorption).
\item \emph{LO-phonon emission} ($\hbar\omega_{\rm LO}=36.6$\,meV, $\tau_{\rm LO}=0.15$\,ps) occurs for the fraction $f$ of electrons whose thermal plus coherent energy exceeds $\hbar\omega_{\rm LO}$. Each event randomizes the momentum, absorbing $\simeq2E_k$ from the field, and deposits $\hbar\omega_{\rm LO}$ in the lattice. The net effect, $(f/\tau_{\rm LO})(2E_k-\hbar\omega_{\rm LO})$, cools the electrons below threshold and heats them above it.
\item \emph{Acoustic-phonon cooling} ($\tau_{\rm ac}=1$\,ns) is orders of magnitude weaker than the drive heating at $T_e\approx30$--40\,K (about $4\times10^{-11}$\,W per electron at $U_p=6$\,meV), which is why the results are insensitive to $\tau_{\rm ac}$.
\end{itemize}

\textit{Self-consistent CW solution.}---For CW driving, the electron temperature and the cavity response are coupled: $T_e$ depends on the quiver energy, while the cavity frequency, losses and noise depend on $T_e$. In the drive-frame model used for Figs.~3--6 and Table~II of the main text, we parametrize the steady state by the dressed photon number $n_c$, or equivalently $U_p$, which makes the self-consistency explicit and avoids iteration. For a given $U_p$, we solve the cycle-averaged energy balance
\begin{equation}
2\gamma U_p+\frac{1}{\tau_{\rm LO}}\Big\langle f(E_k,T_e)\,\big(2E_k-\hbar\omega_{\rm LO}\big)\Big\rangle_\phi-\frac{U(T_e)-U(T_L)}{\tau_{\rm ac}}=0
\end{equation}
for $T_e$, using Brent's method. Here $E_k=2U_p\cos^2\phi$ is the instantaneous coherent kinetic energy at drive phase $\phi$, and $f(E_k,T_e)=\exp\{-\max[\hbar\omega_{\rm LO}-U(T_e)-E_k,0]/k_BT_e\}$ is the fraction of electrons above the LO-phonon threshold. The resulting $T_e$ fixes $\gamma_{\rm eff}=\gamma+\langle f\rangle_\phi/\tau_{\rm LO}$, the absorptive loss $\kappa_{\rm abs}=\gamma_{\rm eff}(R^2-1)/R^2$, the thermal occupation $n_{\rm th}(\omega'_{\rm cav},T_e)$, and the shift of $\omega'_{\rm cav}$ through $D_{\rm eff}(T_e)$. All of these enter the drift and noise matrices of the linearized fluctuations. The drive detuning is then optimized for minimum output noise, and the drive amplitude required to sustain $n_c$ follows from the mean-field relation $|\varepsilon|^2=n_c[(\kappa/2)^2+(\Delta_0+\chi_{\rm K}n_c)^2]$ (End Matter F of the main text), with the heating-modified $\kappa$ and $\Delta_0$. Points at which the linearized dynamics is unstable are discarded. In the non-RWA time-domain model, the same coupling is solved by iteration instead: starting from $T_e=T_L$, the mean field and covariance are computed at fixed $T_e$, the resulting $U_p$ gives an updated $T_e$, and $T_e$ is relaxed with 50\% mixing until it changes by less than 0.5\,K.

\textit{Regimes.}---Below threshold, LO emission by the thermal tail balances the Drude heating at $T_e\approx25$--40\,K, reached within 1--2\,ps. We define the \emph{runaway threshold} as the drive at which the peak coherent kinetic energy plus the mean thermal energy reaches the LO-phonon energy, $2U_p+\langle\varepsilon\rangle_{T_e}\simeq\hbar\omega_{\rm LO}$, i.e., $U_p\approx(\hbar\omega_{\rm LO}-\langle\varepsilon\rangle)/2\approx10$--15\,meV ($n_c\approx10^6$; gray bands in Figs.~2 and 3 of the main text). Beyond it the LO term becomes net heating and $T_e$ runs away, with no stable low-temperature solution (see \textit{Heating limit} in the main text). Because $T_e$ equilibrates within picoseconds, much faster than the cavity build-up time ($1/\kappa\approx7$--18\,ps for $Q=40$--100), pulses cannot avoid heating. Narrowband driving is nevertheless far less harmful than broadband single-cycle driving at equal fluence, since a single-cycle pulse concentrates the energy into a peak field about 13 times higher, which exceeds the runaway threshold.

\textit{Feedback on the cavity.}---$T_e$ enters through $D_{\rm eff}(T_e)$, which shifts $\omega'_{\rm cav}$ (compensated by retuning $\omega_d$); through the absorptive loss $\kappa_{\rm abs}=\gamma_{\rm eff}(R^2-1)/R^2$, with $(R^2-1)/R^2$ the matter fraction of the dressed mode and $\gamma_{\rm eff}=\gamma+f/\tau_{\rm LO}$; and through the thermal occupation $n_{\rm th}(\omega'_{\rm cav},T_e)$ of that channel. The collection efficiency is $\eta=\kappa_{\rm rad}/(\kappa_{\rm rad}+\kappa_{\rm abs})$.

\textit{Figure of merit.}---At the bistability onset the Kerr shift equals $\tfrac{\sqrt3}{2}\kappa$. With $|\chi_{\rm K}|n_c=6|\alpha_{\rm np}|DU_p/R$ this gives $\kappa_{\rm bis}\simeq6\sqrt3|\alpha_{\rm np}|DU_p/R$. Comparing with $\kappa_{\rm abs}=4\gamma D/(\omega_{\rm cav}R^2)$ at $U_p=U_{\max}$ yields Eq.~(3) of the main text. The mode volume, $N$ and $D$ drop out except through $R$, and a smaller $D$ must be matched by a proportionally smaller $\kappa$, giving $Q_{\rm opt}\approx20\,\omega_{\rm cav}/D$.

\textit{Limitations.}---The model uses a single $T_e$, a Boltzmann tail for $f$, a uniform field, a constant $K$, and a temperature-independent $\gamma$. Acoustic-phonon scattering at $T_e\approx30$--40\,K and any difference between the THz and DC scattering rates would increase $\gamma$, which is equivalent to lowering $\mu$ (Fig.~5 and Table~II of the main text). Intervalley transfer, important at the fields of Ref.~\cite{Kim2026}, plays no role below the runaway threshold.

\section{Nanoslot sample: non-radiative damping}\label{sec:nanoslot}

End Matter E of the main text shows that the nanoslot resonator of Ref.~\cite{Kim2026} has a large non-radiative damping, which explains its early heating and the absence of squeezing in that structure. Here we list the inputs to that analysis and how it could be tested. The analysis uses three quantities measured or simulated for the 450-nm nanoslot:
\begin{enumerate}
\item the dressed FWHM at $B=0$ and weak field (0.42\,THz);
\item the effectively bare FWHM at 7\,T (0.22\,THz);
\item 
the absorption fraction $A_0\approx0.6$ (energy absorbed in the quantum well divided by the incident energy), obtained from electromagnetic simulations of the structure in Ref.~\cite{Kim2026}, which include the Drude response of the 2DEG ($\tau=m^*\mu/e$) and the enhanced gap field.
\end{enumerate}
The narrow gap was chosen for a separate study of nonlocal effects, not to minimize absorption, so the large non-radiative damping is a property of that design rather than of the USC platform. A gap-width series (from $\sim$2\,$\mu$m down to a few nm, as fabricated for the nonlocality study) would separate the gap-dependent non-radiative width, which should grow as the gap narrows, from the radiative coupling. The extrapolation to wide gaps would then test the bath-coupling predictions of Table~\ref{tab:bath} directly. For weakly confined cavities (photonic crystals, Fabry--Perot cavities, or antenna lattices with $h\gtrsim1\,\mu$m), $qv_F/\omega\approx35\,{\rm nm}/h\ll1$ and nonlocal damping is negligible.

\section{Operating points in SI units}\label{sec:ops}

Table~\ref{tab:SI} gives incident fields for points A--C of Table~II, assuming radiation-limited input coupling; the field enhancement scales as $Q^{1/2}$ relative to the nanoslot, whose local-field enhancement ($\times40$) was calibrated in Ref.~\cite{Kim2026}. The drive amplitude $\Omega_0/\omega'_{\rm cav}$ is the robust quantity; the incident field depends on how the specific cavity is coupled. Multi-cycle pulses with $\tau\gtrsim5/\kappa$ and peak fields 1.1--1.2 times the CW values give equivalent results.

\begin{table}[h]
\caption{Incident CW fields at the heating-limited optima ($D/\omega_{\rm cav}=0.22$, 0.87\,THz).}
\label{tab:SI}
\begin{ruledtabular}
\begin{tabular}{cccccc}
Point & $Q$ & $\mu$ ($10^6$\,cm$^2$/Vs) & $\Omega_0/\omega'_{\rm cav}$ & $\mathcal E_0$ (V/cm) & intensity (W/cm$^2$)\\ \hline
A & 40 & 1.2 & 18 & 29 & 1.1\\
B & 100 & 1.2 & 8.6 & 22 & 0.63\\
C & 100 & 2.0 & 8.1 & 20 & 0.55\\
\end{tabular}
\end{ruledtabular}
\end{table}

\section{Proposed measurement protocol}\label{sec:measure}

\textit{Classical precursors.}---Before attempting quantum measurements, the parameters of the linearized model can be determined classically (see End Matter F of the main text):
\begin{enumerate}
\item Kerr bistability and hysteresis under multi-cycle drives of increasing amplitude, which calibrate $\chi_{\rm K}$.
\item Phase-sensitive parametric gain: a weak probe at $\omega_d+\delta$ and the generated idler at $\omega_d-\delta$, which give $|g|/\kappa$ and $\Delta/\kappa$, and hence the predicted $N_\theta(\Omega)$ with no free parameters.
\item The ratio of dressed to bare linewidth in a weakly confined cavity (Table~\ref{tab:bath}).
\item Electron thermometry via the drive-dependent redshift, which calibrates the heating threshold.
\end{enumerate}

\textit{Quantum measurement.}---Sub-cycle electro-optic sampling (EOS) measures the field at a chosen phase of the drive and has resolved THz vacuum fluctuations and squeezed vacua~\cite{Riek2015,Riek2017,BeneaChelmus2019}. With a fixed sampling phase relative to the drive, the field variance oscillates at $2\omega_d$; two-probe correlations give $N_\theta(\Omega)$ directly.

\textit{Phase tolerance and efficiency.}---At the heating-limited optimum B, the output squeezing at $\omega_d$ exceeds 2\,dB within $\pm5^\circ$ of $\theta_0$, exceeds 1\,dB within $\pm10^\circ$, and vanishes beyond $\approx\pm15^\circ$. Detection with efficiency $\eta_{\rm det}$ maps $N\to1-\eta_{\rm det}(1-N)$, so the collection efficiency $\eta$ of Table~II and $\eta_{\rm det}$ act multiplicatively. The required phase stability ($\lesssim0.1$\,rad) is compatible with phase-locked THz generation and EOS.

\section{Outlook: $B\neq0$ and beyond the (nonlinear) Hopfield model}\label{sec:outlook}

At $B\neq0$ the cyclotron mode becomes a dynamical degree of freedom with its own bath, and the Kane Landau ladder adds anharmonicity to the matter oscillator. Squeezing then requires a two-mode treatment, with two polariton branches, cavity--cyclotron two-mode squeezing, and a Floquet generalized master equation for the dissipators. Two effects may help: (i) resonant enhancement near the anticrossing, and
(ii) the suppression of momentum relaxation near integer filling (quantum-Hall regime), which could reduce $\gamma$ and hence the heating. The LO-phonon threshold and the $1/N$ dilution are unchanged. Beyond the bosonic Hopfield description, time-dependent Holstein--Primakoff fluctuations about the instantaneous Bloch vector would capture matter squeezing and cross-Kerr effects.

\end{document}